\documentclass[manuscript]{aastex}

\usepackage{amsmath}
\usepackage{amssymb} 

\usepackage{cool} 
\usepackage{breqn} 
\usepackage{bm} 

\usepackage{graphicx,natbib,url,twoopt}
\usepackage{hyperref}                

\hypersetup{
  colorlinks=true,   
  citecolor=blue,
  urlcolor=blue,     
  linkcolor=red,     
}
\usepackage{xcolor}
\usepackage{stmaryrd}

\shorttitle{Resonant Absorption of Axisymmetric Modes}
\shortauthors{Giagkiozis et al.}

\DeclareRobustCommand{\fref}[1]{Figure~\ref{#1}}
\DeclareRobustCommand{\eref}[1]{Eq.~\eqref{#1}}

\DeclareRobustCommand{\sref}[1]{Section~\ref{#1}}

\DeclareRobustCommand{\kummerA}{\ensuremath{a}}
\DeclareRobustCommand{\kummerB}{\ensuremath{b}}
\newcommand{\KummerM}[3]{\ensuremath{M\!\!\left(#1,#2; #3 \right)}}

\DeclareRobustCommand{\kummerVar}{\ensuremath{s}}
\DeclareRobustCommand{\mee}{\kappa}

\DeclareRobustCommand{\pp}{\ensuremath{p^{\prime}}}

\DeclareRobustCommand{\Xip}{\ensuremath{\bm{\xi}}}
\DeclareRobustCommand{\Xirp}{\ensuremath{\xi_r}}
\DeclareRobustCommand{\Xiperp}{\ensuremath{\xi_{\perp}}}
\DeclareRobustCommand{\Xipara}{\ensuremath{\xi_{\parallel}}}
\DeclareRobustCommand{\Xiphip}{\ensuremath{\xi_{\varphi}}}
\DeclareRobustCommand{\Xizp}{\ensuremath{\xi_z}}
\DeclareRobustCommand{\Xirip}{\ensuremath{\xi_{r i}}}
\DeclareRobustCommand{\Xirep}{\ensuremath{\xi_{r e}}}
\DeclareRobustCommand{\Bp}{\ensuremath{\bm{B}^{\prime}}}

\DeclareRobustCommand{\Bphii}{\ensuremath{B_{\varphi i}}}
\DeclareRobustCommand{\Bphie}{\ensuremath{B_{\varphi e}}}

\DeclareRobustCommand{\pTp}{\ensuremath{p_T^{\prime}}}
\DeclareRobustCommand{\pTip}{\ensuremath{p_{T i}^{\prime}}}
\DeclareRobustCommand{\pTep}{\ensuremath{p_{T e}^{\prime}}}

\DeclareRobustCommand{\Beq}{\ensuremath{\bm{B}}}

\DeclareRobustCommand{\Beqr}{\ensuremath{B_r}}
\DeclareRobustCommand{\Beqz}{\ensuremath{B_z}}
\DeclareRobustCommand{\Beqphi}{\ensuremath{B_{\varphi}}}

\DeclareRobustCommand{\Beqzi}{\ensuremath{B_{z i}}}
\DeclareRobustCommand{\Beqphii}{\ensuremath{B_{\varphi i}}}

\DeclareRobustCommand{\Beqze}{\ensuremath{B_{z e}}}
\DeclareRobustCommand{\Beqphie}{\ensuremath{B_{\varphi e}}}
\DeclareRobustCommand{\peq}{\ensuremath{p}}
\DeclareRobustCommand{\peqi}{\ensuremath{p_{i}}}

\DeclareRobustCommand{\rhoeq}{\ensuremath{\rho}}
\DeclareRobustCommand{\rhoi}{\ensuremath{\rho_i}}
\DeclareRobustCommand{\rhoe}{\ensuremath{\rho_e}}

\DeclareRobustCommand{\kwvV}{\ensuremath{\mathbf{k}}}
\DeclareRobustCommand{\kwvTh}{\ensuremath{m}}
\DeclareRobustCommand{\kwvZ}{\ensuremath{k_{z}}}
\DeclareRobustCommand{\kwvR}{\ensuremath{k_r}}
\DeclareRobustCommand{\kwvRin}{\ensuremath{k_{r i}}}
\DeclareRobustCommand{\kwvRout}{\ensuremath{k_{r e}}}

\DeclareRobustCommand{\Ni}{\ensuremath{n_i}}

\DeclareRobustCommand{\Ne}{\ensuremath{n_e}}

\DeclareRobustCommand{\omegaA}{\ensuremath{\omega_{A}}}

\DeclareRobustCommand{\omegaAe}{\ensuremath{\omega_{A e}}}
\DeclareRobustCommand{\omegaC}{\ensuremath{\omega_{c}}}

\DeclareRobustCommand{\vA}{\ensuremath{v_{A}}}

\DeclareRobustCommand{\vAi}{\ensuremath{v_{A i}}}

\DeclareRobustCommand{\vAe}{\ensuremath{v_{A e}}}
\DeclareRobustCommand{\vS}{\ensuremath{v_{s}}}

\DeclareRobustCommand{\vSi}{\ensuremath{v_{s i}}}

\DeclareRobustCommand{\vT}{\ensuremath{v_{T}}}

\newcommand{\twopartdef}[4]
{
	\left\{
		\begin{array}{ll}
			#1 & \mbox{for } #2 \\
			#3 & \mbox{for } #4
		\end{array}
	\right.
}

\newcommand{\twopartdefc}[4]
{
	\left\{
		\begin{array}{ll}
			#1 & #2 \\
			#3 & #4
		\end{array}
	\right.
}

\newcommand{\threepartdef}[6]
{
	\left\{
		\begin{array}{ll}
			#1 & \mbox{for } #2 \\
			#3 & \mbox{for } #4 \\
			#5 & \mbox{for } #6
		\end{array}
	\right.
}

\begin{document}


\title{Resonant Absorption of Axisymmetric Modes \\
in Twisted Magnetic Flux Tubes}


\author{I. Giagkiozis\altaffilmark{1}, M. Goossens\altaffilmark{2}, G. Verth\altaffilmark{1}, V. Fedun\altaffilmark{3}, T. Van Doorsselaere\altaffilmark{2}}

\altaffiltext{1}{Solar Plasma Physics Research Centre, School of Mathematics and Statistics, University of Sheffield, Hounsfield Road, Hicks Building, Sheffield, S3 7RH, UK}
\altaffiltext{2}{Centre for mathematical Plasma Astrophysics, Mathematics Department, KU Leuven, Celestijnenlaan 200B bus 2400, B-3001 Leuven, Belgium}
\altaffiltext{3}{Department of Automatic Control and Systems Engineering, University of Sheffield, Mappin Street, Amy Johnson Building, Sheffield, S1 3JD, UK}


\begin{abstract}
It has been shown recently that magnetic twist and axisymmetric MHD modes are ubiquitous in the solar atmosphere and therefore, the study of resonant absorption for these modes have become a pressing issue as it can have important consequences for heating magnetic flux tubes in the solar atmosphere and the observed damping. In this investigation, for the first time, we calculate the damping rate for axisymmetric MHD waves in weakly twisted magnetic flux tubes. Our aim is to investigate the impact of resonant damping of these modes for solar atmospheric conditions. This analytical study is based on an idealized configuration of a straight magnetic flux tube with a weak magnetic twist inside as well as outside the tube. By implementing the conservation laws derived by \cite{Sakurai:1991aa} and the analytic solutions for weakly twisted flux tubes obtained recently by \cite{Giagkiozis:2015apj}, we derive a dispersion relation for resonantly damped axisymmetric modes in the spectrum of the Alfv\'{e}n continuum. We also obtain an insightful analytical expression for the damping rate in the long wavelength limit. Furthermore, it shown that both the longitudinal magnetic field and the density, which are allowed to vary continuously in the inhomogeneous layer, have a significant impact on the damping time. Given the conditions in the solar atmosphere, resonantly damped axisymmetric modes are highly likely to be ubiquitous and play an important role in energy dissipation. 

We also suggest that given the character of these waves, it is likely that they have already been observed in the guise of Alfv\'{e}n waves.
\end{abstract}


\keywords{axisymmetric modes, mhd, resonant absorption}



\section{Introduction}\label{sec:introduction}
Inhomogeneities, such as a density variation across a magnetic flux tube, produce a continuous spectrum of eigenfrequencies. For instance, consider a straight magnetic flux tube of radius $r_e$ and constant temperature, where the density varies smoothly from its center to its boundary, such that cylindrical surfaces have constant density. This means that also the sound and Alfv\'{e}n speeds within every cylindrical surface are constant. These concentric cylindrical sheaths comprise the flux tube. Due to the difference in characteristic speeds, every surface will have its own eigenfrequency. This results in an infinite set of eigenfrequencies, a continuum. One of the consequences of this continuum in driven systems is resonant absorption, assuming the driving frequency is within the continuum. 

Given that inhomogeneities are the rule rather than the exception in the solar atmosphere, resonant absorption is bound to occur there. This has long been recognized, from the first suggestion by \cite{Ionson:1978aa} to subsequent studies motivated by advances in solar observations, see for example the following works \citep{Poedts:1989aa,Poedts:1990aa,ruderman2002damping,goossens2002coronal,andries2005coronal,Goossens:2009aa,van2009effect,verth2010observational,Terradas:2010aa,Antolin:2015aa,Okamoto:2015aa} to name but a few. In general, resonant absorption in magnetohydrodynamic (MHD) modes is important for the solar atmosphere. Some of the many reasons for this are the following. Resonant damping of Alfv\'{e}n waves is a natural and efficient mechanism for energy dissipation of MHD waves in inhomogeneous plasmas \citep{Ionson:1978aa,Ionson:1985aa,Hollweg:1988ab}. It can also provide an explanation for the observed loss of power of acoustic modes in sunspots \citep{Hollweg:1988aa,Sakurai:1991aa,Sakurai:1991ab,Goossens:1992aa,Keppens:1994aa}, and, it has been shown that it is of importance in transverse oscillations (kink mode), see for example \citep{Aschwanden:1999aa,Nakariakov:1999aa,ruderman2002damping,goossens2002coronal}. Resonant Alfv\'{e}n waves can be an energy conduit between photospheric motions at the footpoints of coronal loops \citep[see for example][]{De-Groof:2000aa,De-Groof:2002aa,De-Groof:2002ab}, and, resonant dissipation plays an important role in the observed damped oscillations in prominences \citep[see][]{Terradas:2008aa,Arregui:2012aa}. For an in depth review of resonant absorption in the solar atmosphere see \cite{goossens2011resonant}.

Since 1999, when the first post-flare standing mode transverse oscillations were detected using the Transition Region and Coronal Explorer (TRACE) \citep{Aschwanden:1999aa,Nakariakov:1999aa} there has been a growth in studies of resonant absorption for the kink mode. \cite{ruderman2002damping} produced relations describing the expected damping for coronal loops using the long wavelength and pressure-less plasma\footnote{Also referred to as \textit{cold} plasma approximation.} approximations, a result that was previously obtained by \cite{goossens1992resonant} using the connection formulae derived by \cite{Sakurai:1991aa,Sakurai:1991ab} for the driven problem and by \cite{tirry1996quasi} for the eigenvalue problem. Later \cite{goossens2002coronal} and \cite{Aschwanden:2003aa} used these results and calculated the expected damping times for a sequence of observed parameters for coronal flux tubes. \cite{goossens2002coronal} concluded that for the parameter sample used, resonant absorption can explain the observed damping times well, provided that the density contrast is allowed to vary from loop to loop. Another important result in this work is that the observed damping does not require modification of the order of magnitude estimates of the Reynolds number ($10^{14}$) as suggested by \cite{Nakariakov:1999aa}. \cite{Aschwanden:2003aa} also arrived at the conclusion that, on average, the theoretical predictions of the damping rate derived by \cite{goossens1992resonant} and \cite{ruderman2002damping}, are consistent with observations and suggested that damping times of coronal loops can be used to infer their density contrast with the surrounding plasma. Coronal flux tubes tend to deform in their \textit{middle} section due to buoyancy, effectively resulting in cross-sections that are approximately elliptical. \cite{ruderman2003resonant} studied the damping of the kink mode in flux tubes with an elliptical cross-section and found that for moderate ratios of the minor to major semi-axis the difference of the damping rate for resonant absorption compared with flux tubes with circular cross-section is not very large. Another deviation from the ideal straight magnetic flux tube is axial curvature. \cite{van2004effect} studied the effect of this curvature and also found that the longitudinal curvature of flux tubes does not significantly alter the damping time of kink modes. Progressively the theoretical models for kink oscillations have become more elaborate, for example, \cite{andries2005coronal} considered longitudinal density stratification. Also, methods for kink wave excitation have been studied, see for example \cite{Terradas:2009aa}. The increased body of observations of kink waves allowed \cite{Verwichte:2013ab} to perform a statistical study to constrain the free parameters present in theoretical models of resonant absorption in kink modes. 

In contrast to this avalanche of theoretical and observational advances related to the kink mode, resonant absorption for axisymmetric modes has not received much attention. One reason for this is that it was believed that the sausage mode had a long wavelength cutoff \citep[e.g.][]{edwin1983wave} which suggested that observation of the sausage mode would be quite challenging. Furthermore, it was correctly believed that for a straight magnetic field, axisymmetric modes could not be resonantly damped. However, it is apparent, even in early works in resonant absorption \citep[see for example][]{Sakurai:1991aa,Sakurai:1991ab,goossens1992resonant}, that for weakly twisted magnetic field axisymmetric modes can and are resonantly damped. What was not known until recently, however, was that the long wavelength cutoff for these modes is also removed in the presence of weak magnetic twist \citep{Giagkiozis:2015apj}. Therefore these modes can freely propagate for all wavelengths. And so, at least in principle, these modes should be observable. Additionally, recent works suggest that magnetic twist and axisymmetric modes are ubiquitous throughout the solar atmosphere. Therefore, the study of these modes has become quite relevant and important. Some examples of magnetic twist in the solar atmosphere are, flux tubes emerging from the convection zone \citep[see for example][]{hood2009emergence,luoni2011twisted}, sunspot rotation can result in twisted magnetic fields \citep{brown2003observations,yan2007rapid,kazachenko2009sunspot}, spicules are observed to have twist \citep{de2012ubiquitous,sekse2013interplay} as well as solar tornadoes \citep{wedemeyer2012magnetic}. Lastly observations of axisymmetric modes have been recently reported in \cite{morton2012observations} and \cite{Grant:2015aa}.

In this work, we focus on the resonant absorption of axisymmetric MHD modes in weakly twisted magnetic flux tubes. Axisymmetric modes correspond to modes with azimuthal wavenumber $\kwvTh=0$. We accomplish this using the following sequence. First we recall recent results for axisymmetric modes in magnetic flux tubes with weak twist \citep{Giagkiozis:2015apj}. In that work the longitudinal component of the magnetic field, and the density were discontinuous across the flux tube boundary. This choice was intentional as it avoids the MHD continua and simplifies the analysis. However, this also left out relevant physics. Then having as a starting point the setup in \cite{Giagkiozis:2015apj} we introduce an intermediate layer about the flux tube boundary. Within this layer, we allow the magnetic field and density to vary smoothly, resulting in an overall continuous profile for the longitudinal magnetic field and density. This in turn allows for the existence of the two MHD continua, the slow and Alfv\'{e}n continuum. Next, we assume that the layer that connects the internal and external quantities, is thin, namely we assume that $\ell \ll r_e$ where $\ell$ is the width of the layer and $r_e$ is the flux tube radius. Then we use the conservation laws, and the resulting jump conditions, for the Alfv\'{e}n continuum by \cite{Sakurai:1991aa}, and we derive the resulting complex dispersion relation. We then solve this dispersion relation numerically. Lastly, to better understand the predicted damping times we apply the long wavelength limit approximation to the resulting complex dispersion relation. These simpler relations allow us to compare our results with the expected damping for the kink mode predicted using the results by \cite{goossens1992resonant} and \cite{ruderman2002damping}. We conclude this investigation with a statistical analysis of the resulting approximations to further understand the necessary conditions for the observation of resonantly damped axisymmetric modes. The main contributions of this work can be summarized as follows. 
\begin{itemize}
\item For the first time, we uncover a dispersion relation for axisymmetric modes in magnetic flux tubes with internal and external twist, including the resonance with the Alfv\'{e}n continuum. We produce simplified expressions for the frequency and damping time in the long wavelength limit, for which the axisymmetric modes are no longer leaky.
\item Given that there are four parameters required for the evaluation of the aforementioned relation, namely density contrast, magnetic field contrast, thickness of the inhomogeneous layer and magnetic twist, we present a statistical framework to infer what can be drawn from observations. 
\item We use this statistical framework and show that the predictions of our theoretical model are in agreement with observed damping times that are in agreement with observed damping times of quasi periodic pulsations (QPPs). QPPs are interpreted as axisymmetric modes (sausage modes) \citep{Kolotkov:2015aa}. 
\end{itemize}

The plan of this paper is as follows. In \sref{sec:model} we present the model, and include prior theoretical results required for the derivation of the dispersion relation leading to resonant absorption. In \sref{sec:alfven:continuum}, using the jump relations in \cite{Sakurai:1991aa} we derive a dispersion equation. In \sref{sec:long} we use the dispersion relation derived in \sref{sec:alfven:continuum} to obtain an expression for the damping rate in the long wavelength limit and then in \sref{sec:con:observations} we elaborate on the significance of the results in this work for the observation of axisymmetric modes in the solar atmosphere. Lastly, in \sref{sec:conclusions} we summarize and conclude this work.

\section{Model}\label{sec:model}
\begin{figure}
\epsscale{0.7}
\plotone{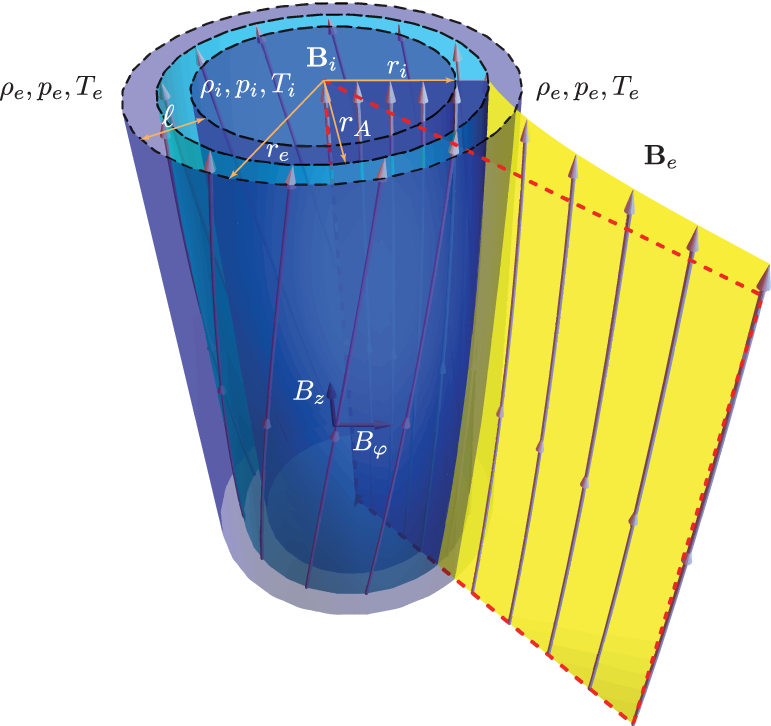}
\caption{Illustration of the model used in this paper. Straight magnetic cylinder with variable twist inside ($r<r_i$) and outside ($r>r_e$) the tube. The region where $r_i < r < r_e$ is the inhomogeneous layer, where the $\Beqz$ component of the magnetic field and the density are varying continuously across this layer. The parameters $\rhoi, \peqi$ and $T_i$ are respectively the density, kinetic pressure and temperature at equilibrium inside the tube, i.e. for $r<r_i$. The corresponding quantities outside the tube ($r>r_e$) are denoted with a subscript $e$. Also, $r_A$ is the radius at the resonance. The dark blue surface emanating radially outwards inside the tube represents the influence of $\Beqphi \propto r$. The yellow surface outside the tube corresponds to $\Beqphi \propto 1/r$ dependence. The dashed red rectangle depicts a magnetic surface which would correspond to a magnetic field with only a longitudinal ($z$) magnetic field component. The inhomogeneous layer is bounded between $r_i$ and $r_e$ and is of width $\ell$. Note that the radius of the tube with the inhomogeneous layer is $r_e$.}\label{fig:tube:twist}
\end{figure}

In this work we assume an idealized cylindrically symmetric magnetic flux tube in static equilibrium. We employ cylindrical coordinates $r, \varphi$ and $z$, with the $z$ coordinate along the axis of symmetry of the flux tube. The linearized ideal MHD equations are, 
\begin{dgroup}\label{eqn:linearized:mhd}
\begin{dmath}
\rhoeq \pderiv[2]{\Xip}{t} + \nabla \pp + \frac{1}{\mu_0}\left( \Bp \times (\nabla \times \Beq) + \Beq \times (\nabla \times \Bp) \right) = 0,\qquad
\end{dmath}
\begin{dmath}
\pp + \Xip \cdot \nabla \peq + \gamma \peq \nabla \cdot \Xip = 0,\qquad
\end{dmath}
\begin{dmath}
\Bp + \nabla \times (\Beq \times \Xip) = 0,\qquad
\end{dmath}
\end{dgroup}
where $\rhoeq, \peq$ and $\Beq$ are the density, plasma kinetic pressure and magnetic field, respectively, at equilibrium, $\Xip$ is the Lagrangian displacement, $\pp$ and $\Bp$ are the Eulerian variations of the pressure and magnetic field, $\gamma$ is the ratio of specific heats (taken to be $5/3$ in this work), and $\mu_0$ is the permeability of free space. In what follows an index, $i$, indicates quantities inside the flux tube ($r<r_i$) while variables indexed by, $e$, refer to the environment outside the flux tube ($r>r_e$). The inhomogeneous layer has a width equal to $\ell = r_e - r_i$ and it is assumed that $\ell \ll r_e$. Note that in \cite{Giagkiozis:2015apj}, $r_a$, was used to denote the tube radius, this is equivalent to $r_e$ in this work. The model configuration is illustrated in \fref{fig:tube:twist} when $\Beqphie \propto 1/r$. The quantities $\rhoeq, \peq$ and $\Beq$ are assumed to have only an $r$-dependence, therefore, the following balance equation must be satisfied when $\ell = 0$, 
\begin{dmath}\label{eqn:pressure:r}
 \D{}{r}\!\left(\peq + \frac{\Beqphi^2 + \Beqz^2}{2 \mu_0} \right) = - \frac{\Beqphi^2}{\mu_0 r}.
\end{dmath}
The equilibrium magnetic field is taken to be $\Beq = (0, \Beqphi, \Beqz)$, with $\Beqphii = S r$, $\Beqphie = r_e^{1+\mee} S / r^{\mee}$ and $\Beqzi,\Beqze$ constant. By substituting $\Beqphii$ and $\Beqphie$ into \eref{eqn:pressure:r} and defining $B_{\varphi A} = \Beqphi(r_e) = S r_e$, we obtain:
\begin{dmath}\label{eqn:pressure:pr}
\displaystyle p(r) = \twopartdef{\displaystyle  \frac{B_{\varphi A}^2}{\mu_0}\left(1 - \frac{r^2}{r_e^2}\right) + p_e } {r \leq r_e,} {\displaystyle  \frac{r_e^{2\mee} B_{\varphi A}^2(1- \mee)}{2 \mu_0 \mee} \left( \frac{1}{r^{2 \mee}} - \frac{1}{r_e^{2\mee}}\right) + p_e } {r>r_e,}
\end{dmath}
where, $p_e$, is the pressure at the boundary of the magnetic flux tube and the parameter $\mee \rightarrow 1$ corresponds to external twist proportional to $1/r$ while $\mee \rightarrow 0$ to constant external twist.  Note that although $p(r)$ is continuous, for solar atmospheric conditions and for weak magnetic twist ($ \sup(\Beqphi^2 / \Beqz^2) \ll 1$) its variation is much smaller than $p_e$ and therefore can be assumed to be constant \citep{Giagkiozis:2015apj}. However, in the model used by \cite{Giagkiozis:2015apj} the equilibrium density and the $z$ component of the magnetic field are discontinuous, therefore the Alfv\'{e}n continuum was avoided. Note that in \cite{Giagkiozis:2015apj} the equivalent to \eref{eqn:pressure:pr} had a typographical error, $(1 - 2 \mee)$ should read $(1-\mee)$.

\begin{figure}
\centering
\plotone{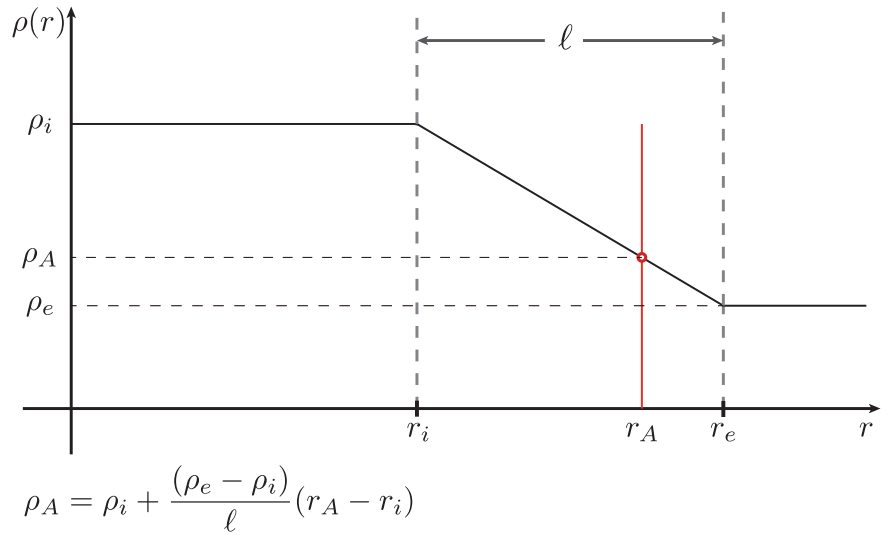}
\caption[]{\label{fig:density:profile}
Density profile as a function of $r$ in the inhomogeneous layer of the magnetic flux tube. Here, $r_i$ and $r_e$ are the radius at which the inhomogeneous begins and ends respectively, also, $r_e$ is the flux tube radius. Lastly, $r_A$, is the radius at the resonance.
}
\end{figure}

In the present investigation both the density and the magnetic field are continuous, see \fref{fig:density:profile}, which introduces the slow and fast continua into our model. Specifically, the density is assumed to be a piecewise linear function of the following form, 
\begin{dmath}\label{eqn:density:profile}
\displaystyle \rho(r) = \threepartdef{\displaystyle \rhoi}{r < r_i,}{\displaystyle \rhoi + \frac{r-r_i}{\ell} (\rhoe - \rhoi) }{r_i \leq r \leq r_e,}{\displaystyle \rhoe}{r > r_e,}
\end{dmath}
a similar form for the variation in the longitudinal component of the magnetic field is assumed, namely, 
\begin{dmath}\label{eqn:bz:profile}
\displaystyle \Beqz(r) = \threepartdef{\displaystyle \Beqzi}{r < r_i,}{\displaystyle \Beqzi + \frac{r-r_i}{\ell} (\Beqze - \Beqzi) }{r_i \leq r \leq r_e,}{\displaystyle \Beqze}{r > r_e.}
\end{dmath}
Note that the assumption here is that $\ell \ll r_e$, so that pressure balance is maintained (see \eref{eqn:pressure:r}). Also note that allowing both the density and the magnetic field to vary results in a non-monotonic variation in the Alfv\'{e}n frequency across the inhomogeneous layer as seen in \fref{fig:omegaA:across:layer}.

The equilibrium quantities depend only on $r$ and therefore the perturbed quantities can be Fourier analyzed with respect to the $\varphi$ and $z$ coordinates, namely, 
\begin{dmath}\label{eqn:fourier:xi:pt}
\Xip, \pTp \propto e^{i\left(\kwvTh \varphi + \kwvZ z - \omega t\right)}.
\end{dmath}
Here, $\omega$ is the angular frequency, $\kwvTh$ is the azimuthal wavenumber, $\kwvZ$ is the longitudinal wavenumber, and $\pTp$ is the Eulerian total pressure perturbation defined as $\pp + \Beq \Bp/\mu_0$. Our focus is on axisymmetric modes (sausage waves) and therefore the azimuthal wavenumber is taken to be $\kwvTh = 0$. The Lagrangian displacement vector in flux coordinates is $\Xip = (\Xirp,\Xiperp,\Xipara)$ where,
\begin{align}\label{eqn:xiperp:xipara:def}
\Xiperp = \frac{\Beqz \Xiphip - \Beqphi \Xizp}{|\Beq|}, && \Xipara = \frac{\Beqphi \Xiphip + \Beqz \Xizp}{|\Beq|},
\end{align}
assuming $\Beqr = 0$. Using \eref{eqn:fourier:xi:pt}, \eref{eqn:linearized:mhd} can be transformed to the following two coupled first order differential equations, 
\begin{dgroup}\label{eqn:sakurai}
\begin{dmath}\label{eqn:sakurai:xidel}
D \D{(r \Xirp)}{r} = C_1 (r \Xirp) - r C_2 \pTp,
\end{dmath}
\begin{dmath}\label{eqn:sakurai:ptdel}
D \D{\pTp}{r} = \frac{1}{r} C_3 (r \Xirp) - C_1 \pTp.
\end{dmath}
\begin{dmath}\label{eqn:xi:flux:perp}
\rhoeq (\omega^2 - \omegaA^2) \Xiperp = \frac{\imath}{|\Beq|} C_A,
\end{dmath}
\begin{dmath}\label{eqn:xi:flux:para}
\rhoeq (\omega^2 - \omega_c^2) \Xipara = \frac{\imath f_B}{|\Beq|}\frac{\vS^2}{\vS^2 + \vA^2} C_S,
\end{dmath}
\begin{dmath}\label{eqn:div:xi}
\nabla \cdot \Xip = - \frac{ \omega^2 C_S }{\rhoeq (\vS^2 + \vA^2)(\omega^2 - \omegaC^2)}
\end{dmath}
\end{dgroup}
and,
\begin{dgroup}\label{eqn:parameters}
\begin{dmath}
D = \rho (\omega^2 - \omega_A^2) C_4,
\end{dmath}
\begin{dmath}
C_1 = \frac{2 \Beqphi}{\mu_0 r} \left( \omega^4 \Beqphi - \frac{\kwvTh}{r} f_B C_4 \right),
\end{dmath}
\begin{dmath}
C_2 = \omega^4 - \left(\kwvZ^2 + \frac{\kwvTh^2}{r^2}\right)C_4,
\end{dmath}
\begin{dmath}
C_3 = \rho D \left[ \omega^2 - \omega_A^2 + \frac{2 \Beqphi}{\mu_0 \rho} \D{}{r}\left( \frac{\Beqphi}{r} \right) \right] \\+ 4 \omega^4\frac{ \Beqphi^4}{\mu_0^2 r^2} - \rho C_4 \frac{ 4 \Beqphi^2 \omega_A^2 }{\mu_0 r^2},
\end{dmath}
\begin{dmath}
C_4 = (\vS^2 + \vA^2)(\omega^2 - \omega_c^2),
\end{dmath}
\begin{dmath}
C_{A} = g_B \pTp - 2 \frac{f_B \Beqphi \Beqz \Xirp}{\mu_0 r}, \; C_S { = } \pTp - 2 \frac{\Beqphi^2 \Xirp}{\mu_0 r}
\end{dmath}
\end{dgroup}
where,
\begin{align*}
\vS^2 &= \gamma \frac{\peq}{\rhoeq}, && \vA^2 = \frac{\Beq^2}{\mu_0 \rhoeq}, \\
\omegaC^2 &= \frac{\vS^2}{\vA^2 + \vS^2} \omegaA^2, && \omegaA^2 = \frac{f_B^2}{\mu_0 \rhoeq}, \\
f_B &= \kwvV \cdot \Beq = \frac{\kwvTh}{r} \Beqphi + \kwvZ \Beqz, && g_B = (\kwvV \times \Beq)_{r} = \frac{\kwvTh}{r} \Beqz
- \kwvZ \Beqphi. \\
\end{align*}
Here, $\kwvV = (0, \kwvTh/r, \kwvZ)$ is the wavevector, $C_A$ and $C_S$ are the coupling functions, $\vS$ is the sound speed, $\vA$ is the Alfv\'{e}n speed, $\omegaC$ is the cusp angular frequency and $\omegaA$ is the Alfv\'{e}n angular frequency. \eref{eqn:sakurai} was initially derived by \cite{hain1958normal} and later by \cite{goedbloed1971stabilization,Sakurai:1991aa}. The first order coupled ODEs in \eref{eqn:sakurai} can be reduced to a single second order ODE for $\Xirp$,
\begin{dmath}\label{eqn:sakurai:ode}
\D{}{r}\left[ \frac{D}{r C_2} \D{}{r}(r \Xirp)\right] + \left[ \frac{1}{D}\left(C_3 - \frac{C_1^2}{C_2} \right) - r \D{}{r}\left( \frac{C_1}{r C_2} \right) \right] \Xirp = 0.
\end{dmath}
The assumption of axisymmetry ($\kwvTh = 0$) leads to, 
\begin{align}\label{eqn:fb:gb:ca:meq0}
f_B = \kwvZ \Beqz, && g_B = -\kwvZ \Beqphi, && C_A = - \kwvZ \Beqphi \left(\pTp + 2 \frac{\Beqz^2}{\mu_0 r} \Xirp \right).
\end{align}
And therefore, 
\begin{dgroup}\label{eqn:xi:flux}
\begin{dmath}\label{eqn:xi:flux:perp:kphi:zero}
\rhoeq (\omega^2 - \omegaA^2) \Xiperp = - \imath \frac{\kwvZ \Beqphi}{| \Beq |} \left( \pTp + 2 \frac{\Beqz^2}{\mu_0 r} \Xirp \right),
\end{dmath}
\begin{dmath}\label{eqn:xi:flux:para:kphi:zero}
\rhoeq (\omega^2 - \omega_c^2) \Xipara = \imath \frac{\kwvZ \Beqz}{| \Beq |} \frac{\vS^2}{\vS^2 + \vA^2}  \left( \pTp - 2 \frac{\Beqphi^2}{\mu_0 r} \Xirp \right).
\end{dmath}
\end{dgroup}
Note that \eref{eqn:xi:flux} suggests that the solutions for the components of the Lagrangian displacement vector are coupled. Coupled in the sense that elimination of one component, e.g. by setting it to be identical to zero, has direct implications to the remaining components. To see this, consider a solution for which $\Xirp = 0$, then by \eref{eqn:sakurai:ode}, $\pTp$ must also be equal to zero and as a consequence of \eref{eqn:xi:flux:perp:kphi:zero} and \eref{eqn:xi:flux:para:kphi:zero} it follows immediately that $\Xiperp$ and $\Xipara$ must also be identically equal to zero. Namely setting $\Xirp = 0$ leads to the trivial solution. Alternatively, let us assume that $\Xiperp = 0$. In this case, by \eref{eqn:xi:flux:perp:kphi:zero} the following relation must hold, 
\begin{dmath}
\pTp = - 2 \frac{\Beqz^2}{\mu_0 r} \Xirp.
\end{dmath}
This in turn implies, 
\begin{dmath}
\rho (\omega^2 - \omega_c^2) \Xipara = - 2 \imath \frac{\kwvZ \Beqz | \Beq |}{\mu_0 r} \frac{\vS^2}{\vS^2 + \vA^2} \Xirp,
\end{dmath}
which in general is non-zero. Now, if we assume $\Xipara = 0$ then,
\begin{dmath}
\pTp = 2 \frac{\Beqphi^2}{\mu_0 r} \Xirp
\end{dmath}
which leads to,
\begin{dmath}
\rho (\omega^2 - \omegaA^2) \Xiperp = -2 \imath \frac{\kwvZ \Beqphi | \Beq |}{\mu_0 r} \Xirp.
\end{dmath}
In the case where $\Beqphi = 0$ then $\Xiperp$ decouples from $\Xirp$ and $\Xipara$. At this point it is instructive to mention the interpretation of the three components of $\Xip$ in flux coordinates by \cite{goossens2011resonant}. \cite{goossens2011resonant} suggest that $\Xiperp$ is the dominant component for Alfv\'{e}n waves and for low plasma-$\beta$ the slow and fast magnetoacoustic waves $\Xipara$ and $\Xirp$ is the dominant component, respectively. A quick check, by setting $\Beqphi = 0$ in \eref{eqn:xiperp:xipara:def}, renders $\Xiperp$ equivalent to $\Xiphip$. This illuminates the connection of $\Xiperp$ with torsional Alfv\'{e}n waves. 

\cite{Giagkiozis:2015apj} solved \eref{eqn:sakurai:ode} for weak internal and external magnetic twist, albeit with the density profile assumed piecewise constant. With the help of the conservation relations for the Alfv\'{e}n continuum derived by \cite{Sakurai:1991aa}, these solutions, which are for ideal MHD, can be used to produce a dispersion relation for MHD waves that undergo damping in the continuum. The solutions by \cite{Giagkiozis:2015apj} are as follows, 
\begin{dgroup}[frame={0pt},framesep=0pt,compact]\label{eqn:solutions:in}
\begin{dmath}\label{eqn:xir:inside:solution}
\Xirip(\kummerVar) = A_{i} \frac{\kummerVar^{1/2}}{E^{1/4}} e^{-\kummerVar/2} \KummerM{\kummerA}{\kummerB}{\kummerVar},
\end{dmath}
\begin{dmath}\label{eqn:pti:inside:solution}
\pTip(\kummerVar) = A_{i} \frac{k_a D_i}{\Ni^2 - \kwvZ^2} e^{-\kummerVar/2}\left[ \frac{\Ni+\kwvZ}{\kwvZ} \kummerVar \KummerM{\kummerA}{\kummerB}{\kummerVar} \\ - 2 \KummerM{\kummerA}{\kummerB-1}{\kummerVar} \right],
\end{dmath}
\end{dgroup}
and,
\begin{dgroup}[frame={0pt},framesep=0pt,compact]\label{eqn:solutions:out}
\begin{dmath}\label{eqn:xir:outside:solution}
\Xirep(r) = A_{e}K_{\nu}\left(\kwvRout r\right),
\end{dmath}
\begin{dmath}\label{eqn:pte:outside:solution}
\pTep =
A_{e} \left(\frac{ \mu_0 (1-\nu) D_e - 2 B_{\varphi A}^2 n_e^2 }{\mu_0 r (\kwvZ^2 - \Ne^2)} K_{\nu}(\kwvRout r) \\ - \frac{ D_e }{\kwvRout} K_{\nu-1}(\kwvRout r) \right).
\end{dmath}
\end{dgroup}
$M(\cdot)$ is the Kummer function and $K(\cdot)$ is the modified Bessel function of the second kind \citep{abramowitz2012handbook}. The solutions in \eref{eqn:xir:inside:solution} and \eref{eqn:pti:inside:solution} were initially derived by \cite{erdelyi2007linear}. The parameters in \eref{eqn:solutions:in} and \eref{eqn:solutions:out} are, 
\begin{align}
a &= 1 + \frac{\kwvRin^2}{4 \kwvZ^2 E^{1/2}}, &&  b = 2, \\
k_a &= \kwvZ(1 - \alpha^2)^{1/2}, && \alpha^2 = \frac{ 4 B_{\varphi A}^2 \omega_{Ai}^2}{\mu_0 r_e^2 \rhoi (\omega^2 - \omega_{Ai}^2)^2}, \\
s &= k_a^2 E^{1/2} r^2, && E = \frac{ 4 B_{\varphi A}^4 n_i^2}{\mu_0^2 r_e^4 D_i^2 \kwvZ^2 (1 - \alpha^2)^2}, \\
\kwvR^2 &= \kwvZ^2\left( 1 - \frac{n^2}{\kwvZ^2}\right), && \kwvR^2 = \frac{(\kwvZ^2 \vS^2 - \omega^2)(\kwvZ^2 \vA^2 - \omega^2)}{(\vA^2 + \vS^2)(\kwvZ^2 \vT^2 - \omega^2)},\\
n^2 &= \kwvZ^2 \frac{ \omega^4 }{ (\omega_{s}^2 + \omega_{A}^2)(\omega^2 - \omegaC^2) }, && \vT^2 = \frac{\vA^2 \vS^2}{\vA^2 + \vS^2}, \\
D_i &= \rhoi (\omega^2 - \omega_{Ai}^2), && D_e = \rhoe (\omega^2 - \omega_{Ae}^2).
\end{align}
and $\nu$ is,
\begin{dmath}\label{eqn:nu:square:twist}
\nu^2(\mee; r) = 1 + 2\frac{r_e^{2\mee} B_{\varphi A}^2}{\mu_0^2 D_e^2 r^{2\mee}}\left\{ 2\frac{r_e^{2\mee} B_{\varphi A}^2 \Ne^2 \kwvZ^2}{r^{2 \mee}} + \mu_0 \rhoe \left[\omegaAe^2(n_e^2(3+\mee)\\-\kwvZ^2(1-\mee)) - (\Ne^2 + \kwvZ^2)(1+\mee)\omega^2 \right]\right\}.
\end{dmath}
This function in \cite{Giagkiozis:2015apj} is evaluated for $\mee = 0$, resulting in an exact solution for constant twist outside the flux tube which is also a zero order approximation for the external solution when magnetic twist is proportional to $1/r$: 
\begin{dmath}[frame={0pt},framesep=0pt,compact]
\nu^2(0; r) = 1 + 2\frac{B_{\varphi A}^2}{\mu_0^2 D_e^2}\left\{ 2 B_{\varphi A}^2 \Ne^2 \kwvZ^2 \\+ \mu_0 \rhoe \left[\omegaAe^2(3 n_e^2-\kwvZ^2) - \omega^2(\Ne^2 + \kwvZ^2) \right]\right\}.
\end{dmath}
Using $\nu = \nu(0;r)$, i.e. constant external magnetic twist, results in solutions, namely \eqref{eqn:xir:outside:solution} and \eqref{eqn:pte:outside:solution}, that have approximately $5\%$ root mean squared error when compared with the exact solution corresponding to $\nu  = \nu(1; r)$, that corresponds to external magnetic twist $\sim 1/r$. For more details see \citep{Giagkiozis:2015apj}.

Imposing continuity for the Lagrangian displacement in the radial direction and total pressure continuity across the flux tube, 
\begin{dgroup}
\begin{dmath}\label{eqn:displacement:continuity}
\left.\Xirip \right\vert_{ r = r_e} = \left.\Xirep \right\vert_{ r = r_e },
\end{dmath}
\begin{dmath}\label{eqn:total:pressure:continuity}
\pTip - \left.\frac{\Bphii^2}{\mu_0 r} \Xirip \right\vert_{ r = r_e } = \pTep - \left. \frac{\Bphie^2}{\mu_0 r} \Xirep\right\vert_{ r = r_e },
\end{dmath}
\end{dgroup}
the following dispersion relation was derived, 
\begin{dmath}\label{eqn:dispersion:equation}
\frac{r_e D_e}{\kwvRout} \frac{K_{\nu-1}(\kwvRout r_e)}{K_{\nu}(\kwvRout r_e)} = \rhoi v_{A\varphi i}^2 \left[ \frac{1}{\kwvRin^2}(\Ni + \kwvZ)^2 - \frac{1}{\kwvRout^2}(\Ne^2 + \kwvZ^2) \right] + \frac{(1-\nu)D_e}{\kwvRout^2} - 2 \frac{D_i}{\kwvRin^2}\frac{M(\kummerA,\kummerB-1;\kummerVar)}{M(\kummerA,\kummerB;\kummerVar)},
\end{dmath}
where $v_{A\varphi i}^2 = B_{\varphi A}^2 / \mu_0 \rhoi$ and $\omega_{A\varphi i}^2 = \kwvZ^2 B_{\varphi A}^2 / \mu_0 \rhoi$.

\subsection{Long Wavelength Limit}
\begin{figure}
\centering
\plotone{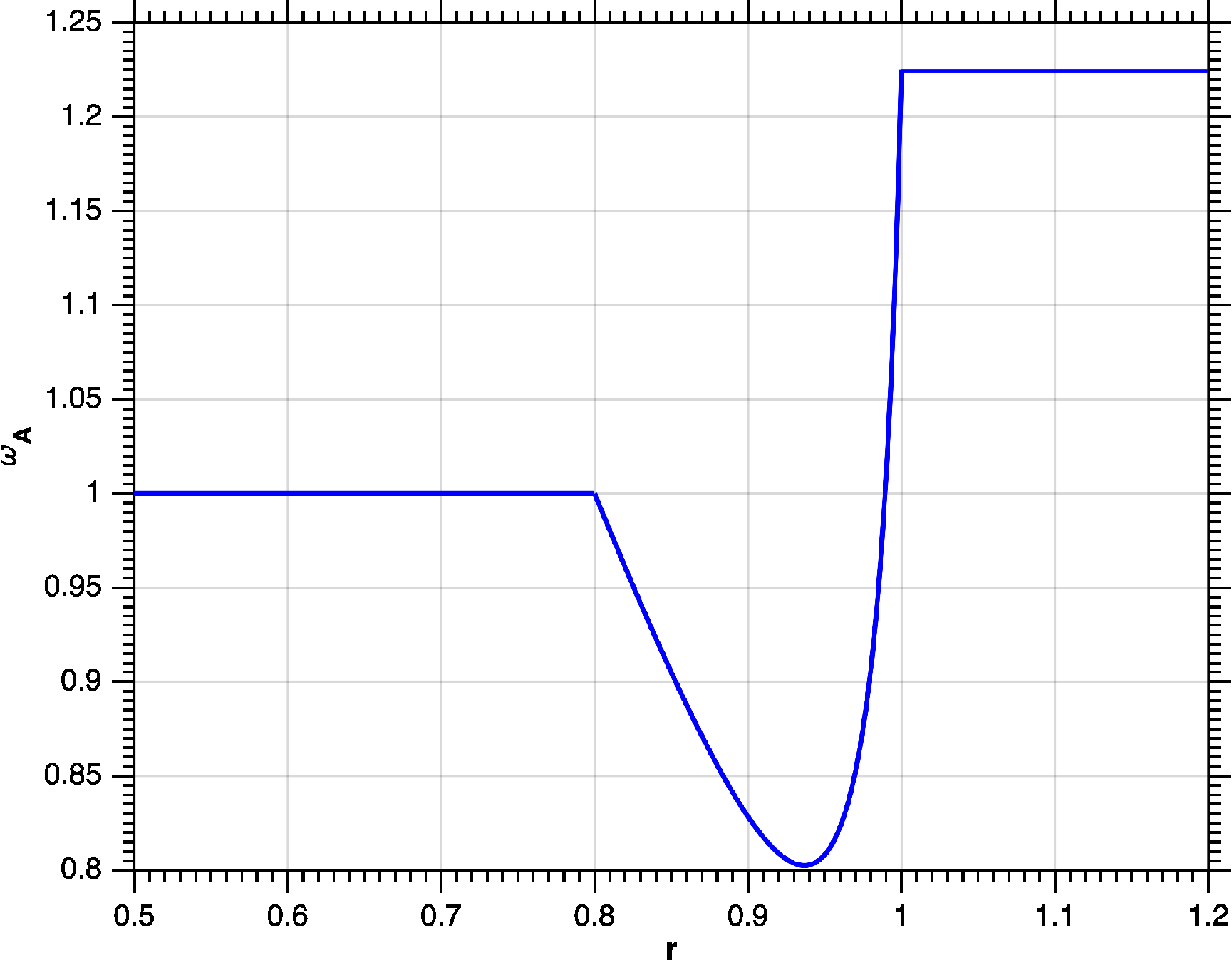}
\caption[]{An example of Alfv\'{e}n frequency variation across the resonant layer when $\Beqz = \Beqz(r)$ and $\rho = \rho(r)$, for $\chi = \rhoe / \rhoi = 0.1$, $\zeta = B_{ze} / B_{zi} = 0.35$ and $\ell / r_e = 0.2$. Here $r=1$ is the tube boundary and $\omega_{A}$ is the normalised normalized Alfv\'{e}n frequency, the normalization is with respect to the internal Alfv\'{e}n frequency, $\omega_{Ai}$.\label{fig:omegaA:across:layer}}
\end{figure}

The long wavelength limit of \eref{eqn:dispersion:equation} is needed for the approximation of the location of the resonant point used in subsequent sections and is obtained as follows. From Eq. (13.5.5) in \cite{abramowitz2012handbook} we have, 
\begin{dmath}\label{eqn:abramo:1}
\underset{\epsilon \rightarrow 0}{\lim} \frac{M(\kummerA,\kummerB-1;\kummerVar)}{M(\kummerA,\kummerB;\kummerVar)} = 1,
\end{dmath}
furthermore, rewriting $\nu^2(0;r)$ as,
\begin{dmath}
\nu^2(0;r) = 1 + 2 \frac{\rhoe B_{\varphi A}^2}{\mu_0 D_e^2}\left( \omega_{Ae}^2\left(3 \frac{\Ne^2}{\kwvZ^2} - 1 \right) - \omega^2\left( \frac{\Ne^2}{\kwvZ^2} + 1 \right) \right) \kwvZ^2 + 4\frac{B_{\varphi A}^4}{\mu_0^2 D_e^2} \frac{\Ne^2}{\kwvZ^2} \kwvZ^4,
\end{dmath}
becomes apparent that $\nu = 1 + \mathcal{O}(\epsilon^2)$, where $\epsilon = r_e \kwvZ$. Therefore using Eq. (9.6.8) and (9.6.9) in \cite{abramowitz2012handbook} we obtain that, 
\begin{dmath}\label{eqn:abramo:2}
\underset{\epsilon \rightarrow 0}{\lim} \frac{K_{0}(\kwvRout r_e)}{K_{1}(\kwvRout r_e)} = 0.
\end{dmath}
Using \eref{eqn:abramo:1} and \ref{eqn:abramo:2} in \eref{eqn:dispersion:equation} we have, 
\begin{dmath}\label{eqn:lwl:twist:1}
2 (\omega^2 - \omega_{Ai}^2) = v_{A\varphi i}^2 \left[ \left( \Ni + \kwvZ \right)^2 - \frac{\kwvRin^2}{\kwvRout^2}\left( \Ne^2 - \kwvZ^2 \right)  \right].
\end{dmath}
Expanding the part in square brackets on the right hand side of this equation about $\epsilon = 0$ leads to,
\begin{dmath}\label{eqn:kz:approximation}
\left( \Ni + \kwvZ \right)^2 - \frac{\kwvRin^2}{\kwvRout^2}\left( \Ne^2 - \kwvZ^2 \right) = 2 \frac{\omega}{(\vAi^2 + \vSi^2)^{1/2}} \kwvZ + \mathcal{O}\left(\epsilon^2\right).
\end{dmath}
Using this approximation in \eref{eqn:lwl:twist:1} the positive solution of the dispersion relation \eref{eqn:dispersion:equation} in the long wavelength limit to first order is,
\begin{dmath}\label{eqn:lwl:twist:first:order}
\omega = \frac{1}{2}\left[ \frac{\omega_{A\varphi i}^2}{(\omega_{Ai}^2 + \omega_{si}^2)^{1/2}} + \left( \frac{\omega_{A\varphi i}^4}{\omega_{Ai}^2 + \omega_{si}^2} + 4 \omega_{Ai}^2 \right)^{1/2}  \right].
\end{dmath}
For notational convenience \eref{eqn:lwl:twist:first:order} is rewritten as follows,
\begin{align}\label{eqn:lwl:twist:first:order:simp}
\omega &= \omega_{Ai} h, \\
h &= \frac{1}{2}\left[ \frac{q_i^2}{\left( 1 + d^2 \right)^{1/2}} + \left( 4 + \frac{q_i^4}{1+d^2} \right)^{1/2}   \right]
\end{align}
where, $q_i = B_{\varphi A} / \Beqzi$ and $ d = \vSi / \vAi$. This $\omega$ is used as an approximation to the resonance frequency, $\omega_0$, in \sref{sec:long}. Lastly, we should note that given this value for $\omega_0$, although the variation of the Alfv\'{e}n speed across the inhomogeneity in the flux tube is quadratic (see \fref{fig:omegaA:across:layer}), since $\omega_{Ai} < \omega_0 < \omega_{Ae}$, there will only be a single resonance point.

\section{Alfv\'{e}n Continuum}\label{sec:alfven:continuum}

For an equilibrium with magnetic twist, such as the model used in this work, the total pressure perturbation is no longer a conserved quantity and therefore \eref{eqn:total:pressure:continuity} and \eref{eqn:displacement:continuity} require modification. \cite{Sakurai:1991aa} derived new conserved quantities for the Alfv\'{e}n and slow continua. Specifically for the Alfv\'{e}n continuum the conserved quantity is, 
\begin{dmath}
C_A = g_{B} \pTp - 2 f_B \Beqphi \Beqz \frac{\Xirp}{\mu_0 r}.
\end{dmath}
Using this conserved quantity they derived \textit{jump conditions} for $\Xirp$ and $\pTp$, namely a prescription on how the radial component of the Lagrangian displacement and the total pressure perturbation can vary across the inhomogeneous layer connecting the internal with the external solutions. This prescription then implies, that the following conditions must be satisfied, 
\begin{dmath}\label{eqn:displacement:continuity:ra}
\left.\Xirip(r)\right\vert_{r = r_i} + \llbracket \Xirp(r) \rrbracket = \left.\Xirep(r)\right\vert_{r=r_e}
\end{dmath}
and
\begin{dmath}\label{eqn:pressure:continuity:ra}
\left.\pTip(r)\right\vert_{r = r_i} + \llbracket \pTp(r) \rrbracket = \left.\pTep(r)\right\vert_{r=r_e},
\end{dmath}
where $\llbracket \Xirp \rrbracket$ and $\llbracket \pTp \rrbracket$ are the jump conditions across the resonant layer in the inhomogeneous section of the flux tube, in the radial displacement and total pressure perturbation \citep{Sakurai:1991aa}. They are given by
\begin{dmath}\label{eqn:displacement:jump}
\llbracket \Xirp \rrbracket = - \frac{ \imath \pi }{ | \Delta_{A} | } \frac{g_B}{ \mu_0 \rho^2 v_A^2} C_A,
\end{dmath}
and 
\begin{dmath}\label{eqn:total:pressure:jump}
\llbracket \pTp \rrbracket = - \frac{ \imath \pi }{ | \Delta_{A} | } \frac{ 2 T \Beqz}{ \mu_0 \rho^2 v_A^2 r} C_A,
\end{dmath}
where $T = f_B \Beqphi / \mu_0 = \kwvZ \Beqz \Beqphi / \mu_0$ and,
\begin{dmath}
\Delta_{A} {=} \D{}{r} \left(\omega^2 - \omega_{A}^2(r) \right).
\end{dmath}
Taking into account that $\kwvTh = 0$ and $\Beqphi \neq 0$ and \eref{eqn:fb:gb:ca:meq0} the jump conditions, \eref{eqn:displacement:jump} and \eqref{eqn:total:pressure:jump}, can be written as, 
\begin{dmath}\label{eqn:displacement:jump:kzero}
\llbracket \Xirp \rrbracket = \frac{ \imath \pi }{ | \Delta_{A} |} \frac{ \kwvZ \Beqphi }{ \mu_0 \rho^2 v_A^2 } C_A = \left. - \frac{ \imath \pi}{| \Delta_{A} | } \frac{ \kwvZ^2 \Beqphi^2}{\mu_0 \rho^2 v_A^2}\right|_{r=r_{A}} \cdot \left. \left\{ \pTp + \frac{ 2 \Beqz^2}{\mu_0}\frac{\Xirp}{r} \right\}\right|_{r=r_i},
\end{dmath}
and,
\begin{dmath}\label{eqn:total:pressure:jump:kzero}
\llbracket \pTp \rrbracket = - \frac{ \imath \pi }{ | \Delta_{A} |} \frac{ 2 \kwvZ \Beqphi \Beqz^2}{ \mu_0^2 \rho^2 v_A^2 r} C_A = \left. \frac{2 \imath \pi}{r | \Delta_{A} |} \left( \frac{\kwvZ \Beqphi \Beqz}{\mu_0 \rho v_A} \right)^{2}\right|_{r=r_A} \cdot \left. \left\{ \pTp + 2 \frac{\Beqz^2}{\mu_0} \frac{\Xirp}{r} \right\} \right|_{r=r_i}.
\end{dmath}
Given that $\Beqz = \Beqz(r)$ and $\rho = \rho(r)$ in the inhomogeneous layer, see \eref{eqn:density:profile}, \eref{eqn:bz:profile} and \fref{fig:density:profile}, we have,
\begin{dmath}\label{eqn:DeltaA}
\Delta_{A} = \omega_{A}(r)^2 \Delta_{AF}
= \omega_{A}(r)^2 \left[ \frac{1}{\rho} \D{\rho}{r} - 2\frac{1}{\Beqz}\D{\Beqz}{r} \right].
\end{dmath}
Obviously when $\Beqz$ is constant across the inhomogeneous layer, 
\begin{dmath}
\Delta_{A} = \omega_{A}^2(r) \frac{1}{\rho} \D{\rho}{r}.
\end{dmath}
Substituting \eref{eqn:displacement:jump:kzero}, \ref{eqn:total:pressure:jump:kzero} and \ref{eqn:DeltaA} into \eref{eqn:displacement:continuity:ra} and \eref{eqn:pressure:continuity:ra} we obtain the 
dispersion relation for axisymmetric MHD waves that undergo resonant absorption in the Alfv\'{e}n continuum of frequencies due to the twist in the magnetic field: 
\begin{dmath}\label{eqn:alfven:continuum:dr}
\mathcal{D}_{AR}(\omega,\kwvZ) + \imath \mathcal{D}_{AI}(\omega,\kwvZ) = 0,
\end{dmath}
where,
\begin{dmath}\label{eqn:alfven:continuum:dr:real}
\mathcal{D}_{AR} = 2 \frac{D_i}{k_{ri}^2} \frac{M(a,b-1;s_i)}{M(a,b;s_i)} - 2 \rho_i n_i (n_i + k_z)\frac{v_{A\varphi i}^2}{k_{ri}^2} 
+ \frac{r_i D_e}{k_{re}} \frac{K_{\nu-1}(k_{re} r_e)}{K_{\nu}(k_{re} r_e)} - \frac{r_i}{r_e k_{re}^2} \left\{  (1 - \nu)D_e - 2 \rho_e n_e^2 v_{A \varphi e}^2 \right\},
\end{dmath}
and,
\begin{dmath}\label{eqn:alfven:continuum:dr:imaginary}
\mathcal{D}_{AI} = \left.\frac{\pi}{\rho |\Delta_{AF}|} \frac{v_{A\varphi}^2}{v_{A}^4} \right|_{r=r_A}\\
 \left[ \frac{2}{k_{ri}^2}
\left(D_i \frac{M(a,b-1;s_i)}{M(a,b;s_i)} - n_i (n_i + k_z) \rho_i v_{A \varphi i}^2 \right) + 2 \rho_i v_{Ai}^2 \right] \\
\left[ 2 \left.\frac{B_z^2}{\mu_0 r}\right|_{r_A} + \frac{1}{r_e k_{re}^2} \left\{ (1-\nu) D_e - 2\rho_e n_e^2 v_{A\varphi e}^2 \right\} 
- \frac{D_e}{k_{re}} \frac{K_{\nu-1}(k_{re} r_e)}{K_{\nu}(k_{re} r_e)} \right].
\end{dmath}
In these equations the following definitions were used, 
\begin{align}
v_{A\varphi A}^2 &= \frac{B_{\varphi A}^2}{\mu_0 \rho_A}, && v_{AA}^2(r_A) = \frac{B_{zA}^2}{\mu_0 \rho_A}, \\ |\Delta_{AF}(r_A)| &= \frac{1}{\ell} \left| 
\frac{\rhoe - \rhoi}{\rho_A} - 2 \frac{\Beqze - \Beqzi}{\Beqz(r_A)} \right|, &&
\end{align}
where $\rho_A = \rho(r_A)$, $v_{AA} = v_{A}(r_A)$ and $B_{zA} = B_{z}(r_A)$.
To find the radius at the resonance point, namely the radius where $v_{A}(r) = v_0 (= \omega_0 / \kwvZ)$\footnote{For the definition of $\omega_0$ see \eref{eqn:lwl:twist:first:order:simp}.}, we can express $r_A$ as a convex combination of the radius $r_i$ and the width of the inhomogeneous layer $\ell$ since $r_A$ must be within the interval $(r_i,r_e)$. Therefore we can write $r_A = r_i + w \ell$, where $w \in (0, 1)$. Now we have transformed the problem of solving for $r_A$ to a problem where we have to solve for $w$, the convex combination parameter. Given this formulation for $r_A$ and \eref{eqn:density:profile} and \eref{eqn:bz:profile} we can write $B_{zA} = B_{z}(r_A) = \Beqzi + w (\Beqze - \Beqzi)$ and $\rho_A = \rho(r_A) = \rhoi + w(\rhoe - \rhoi)$. Equipped with these definitions, the equation that we need to solve to find $w$ becomes,
\begin{dmath}\label{eqn:resonance:radius:0}
v_{AA}^2 {=} \frac{ \left(\Beqzi + w(\Beqze - \Beqzi)  \right)^2 }{\mu_0 (\rhoe + w (\rhoe - \rhoi))  } = v_0^2
\end{dmath}
Using the definitions $\chi = \rhoe / \rhoi$, $\zeta = \Beqze / \Beqzi$ \eref{eqn:resonance:radius:0}, simplifies to
\begin{dmath}\label{eqn:resonance:radius:1}
\vAi^2 \frac{\left( 1 + w (\zeta - 1) \right)^2}{1 + w (\chi - 1)} = v_0^2.
\end{dmath}
This equation is solved for $w$ in the next section. 

\section{Long Wavelength Limit - Alfv\'{e}n Continuum}\label{sec:long}
\begin{figure}
\centering
\epsscale{1}
\plotone{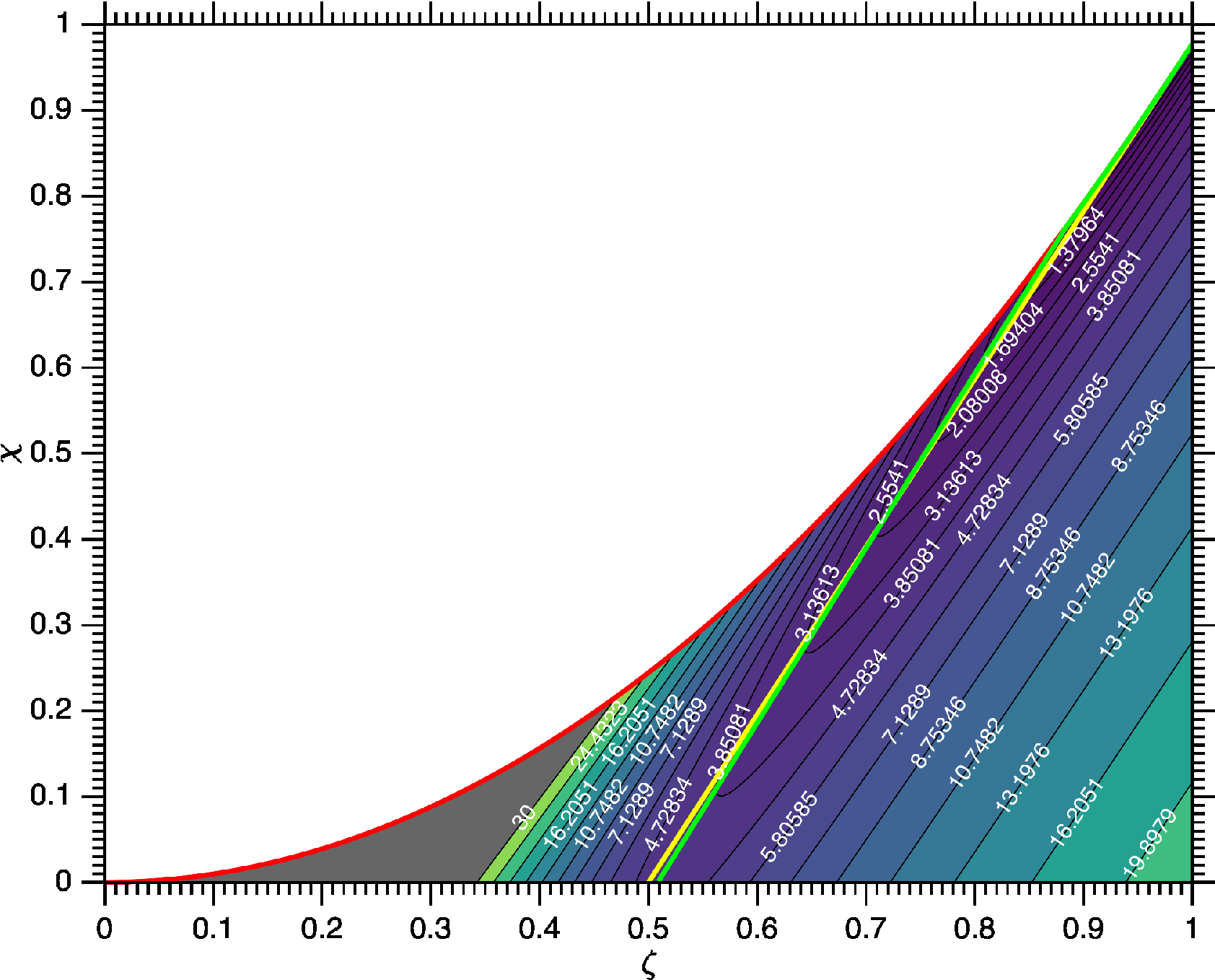}
\caption[]{Contour map of the damping time $\tau_{d}$ (see \eref{eqn:damping:time:bi:d:be}) as a multiple of the period $\tau$, plotted for density contrast in the interval $\chi \in (0,\zeta^2 / h^2)$, and longitudinal magnetic field contrast in the interval $\zeta \in (0,1)$. The remaining parameters in \eqref{eqn:damping:time:bi:d:be} are set as follows: $\ell / r_e = 0.1$ and $B_{\varphi A}/B_{zA} = 0.15$. The \textit{red} line marks $\zeta^2/h^2$ above which the resonance frequency is outside of the continuum. The gray region in this plot denotes damping times of $30$ and above. \label{fig:damping:time:bi:d:be}}
\end{figure}
\begin{figure}
\centering
\epsscale{1}
\plotone{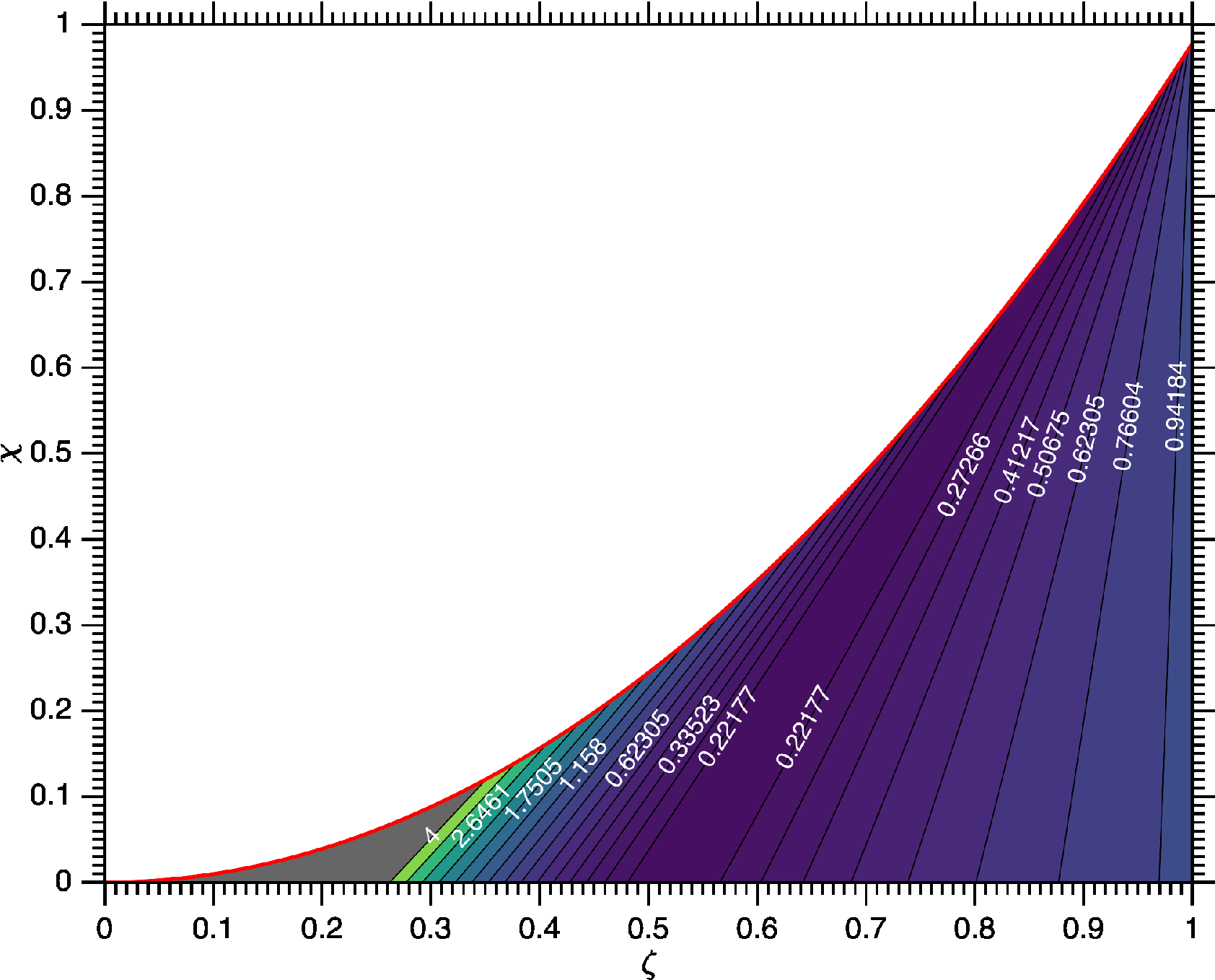}
\caption[]{Contour map of the ratio $\tau_d(\chi,\zeta)$ versus $\tau_d(\chi,1)$, see \eqref{eqn:damping:time:ratio}. The density contrast is allowed to vary in the interval $\chi \in (0,\zeta^2 / h^2)$, and longitudinal magnetic field contrast in the interval $\zeta \in (0,1)$. The \textit{red} line is the same as in \fref{fig:damping:time:bi:d:be}, whilst values within the gray region correspond to ratios larger than $4$. \label{fig:damping:time:ratio}}
\end{figure}
\notetoeditor{\fref{fig:damping:time:bi:d:be} and \fref{fig:damping:time:ratio} should appear side-by-side in the text.}

Taking the long wavelength limit, $\epsilon \ll 1$, of \eref{eqn:alfven:continuum:dr:real} and \eref{eqn:alfven:continuum:dr:imaginary} and using \eref{eqn:abramo:1}, \eref{eqn:abramo:2} then,
\eref{eqn:alfven:continuum:dr:real} and \eref{eqn:alfven:continuum:dr:imaginary} reduce to, 
\begin{dmath}\label{eqn:alfven:continuum:dr:re:lwl1}
\mathcal{D}_{AR} = \frac{\omega^2 - \omega_{Ai}^2}{\kwvRin^2} - \frac{\Ni^2}{\kwvRin^2} v_{A\varphi i}^2 
 + \chi \frac{\Ne^2}{\kwvRout^2} v_{A \varphi e}^2,
\end{dmath}
\begin{dmath}\label{eqn:alfven:continuum:dr:im:lwl1}
\mathcal{D}_{AI} = \frac{\pi}{r_e} \left[\left. \frac{\pi}{\Delta_{AF}} \frac{v_{A\varphi}^2}{v_{A}^2} \right|_{r=r_A}\right] \left[ 1 - \frac{\rhoe}{\rho_A}\frac{\Ne^2}{\kwvRout^2}\frac{v_{A \varphi e}^2}{v_{AA}^2} \right] \\
\left[\frac{\omega^2 - \omega_{Ai}^2}{\kwvRin^2} - \frac{\Ni^2}{\kwvRin^2}v_{A\varphi i}^2 + v_{Ai}^2 \right]
 .
\end{dmath}
These equations can be solved if we allow a complex frequency $\omega = \omega_r + \imath \gamma_A$, and when $\gamma_A \ll \omega_r$ we can obtain the damping rate, $\gamma_A$ in the Alfv\'{e}n continuum frequencies \citep{goossens1992resonant} to second order is given by,
\begin{dmath}\label{eqn:damping:factor:first:order}
\gamma_A = - \mathcal{D}_{AI}(\omega_0)  \left(\left.\frac{\partial\mathcal{D}_{AR}}{\partial \omega}\right|_{\omega=\omega_0}\right)^{-1}.
\end{dmath}
This equation results in an expression that is difficult to interpret, and for this reason, given that we seek an expression for the damping rate in the long wavelength limit we expand it in a series about $\varepsilon = 0$ where $\varepsilon = r_e \kwvZ$. This expansion results in,
\begin{dmath}\label{eqn:damping:factor:bi:d:be}
\gamma_A = \omega_0 \frac{\pi}{Z} \frac{\ell}{r_e} \frac{\rho_A}{\rhoi} \frac{B_{\varphi A}^2}{B_{zA}^2} \left( 1 + \frac{B_{\varphi A}^2}{B_{zA}^2} \right) \left( 1+ \frac{B_{\varphi A}^2}{\Beqzi^2}  \right) + \mathcal{O}\left( \varepsilon^2 \right),
\end{dmath}
where
\begin{dmath}
Z = \left| (\chi-1) - 2 \frac{\rho_A}{\rhoi}\frac{\Beqzi}{B_{zA}} (\zeta - 1) \right|.
\end{dmath}
Now, in this investigation we assume weak magnetic twist ($q = B_{\varphi A} / B_{zA} \ll 1$) and therefore, 
\begin{dmath}
\frac{B_{\varphi A}^2}{B_{zA}^2} \left( 1 + \frac{B_{\varphi A}^2}{B_{zA}^2} \right) \left( 1+ \frac{B_{\varphi A}^2}{\Beqzi^2}  \right) = \frac{B_{\varphi A}^2}{B_{zA}^2} + \mathcal{O}(q^4),
\end{dmath}
and so \eref{eqn:damping:factor:bi:d:be} can be simplified to, 
\begin{dmath}\label{eqn:damping:factor:weak:twist}
\gamma_A = \omega_0 \frac{\pi}{Z} \frac{\ell}{r_e} \frac{\rho_A}{\rhoi} \frac{B_{\varphi A}^2}{B_{zA}^2}.
\end{dmath}
Here $\omega_0$ is approximated by \eqref{eqn:lwl:twist:first:order}, i.e. $\omega_0 \approx \omega_{Ai} h$, and the radius at the resonance point, $r_A$, is obtained analogously to \eref{eqn:resonance:radius:1}, by solving, 
\begin{dmath}\label{eqn:resonance:radius:2}
\frac{\left( 1 + w (\zeta - 1) \right)^2}{1 + w (\chi - 1)} = h^2.
\end{dmath}
There are two cases to be considered. First, when $\zeta = 1$, that is $\Beqzi = \Beqze$, and assuming $\chi \in (0,1/h^2)$, the solution for $w$ is
\begin{dmath}\label{eqn:w:constant:b}
w = \frac{h^2 - 1}{h^2 (1-\chi)},
\end{dmath}
when $1/h^2 < \chi < 1$ in this case the resonant point is outside of the continuum and there is no resonant absorption and in the limit $\chi \rightarrow 1$ the external and internal Alfv\'{e}n speeds become equal and there are no propagating waves either. The second case is for values of $\zeta \in (0,1)$ and $\chi \in (0,\zeta^2 / h^2)$ for which the admissible solution is,
\begin{dmath}\label{eqn:w:varying:b}
w = \frac{  2(1-\zeta) + h\left( 4(\zeta - 1)(\zeta - \chi) + h^2(\chi -1)^2  \right)^{1/2} + h^2 (\chi - 1)  }{ 2\left( 1 - \zeta\right)^2  }.
\end{dmath}
When $\chi > \zeta^2 / h^2$, similarly to the first case there is no resonant absorption since the resonance frequency ($\omega_0$) is outside of the continuum. For $\zeta^2 / h^2 < \chi < \zeta^2$ there exist undamped propagating waves, however, when $\chi > \zeta^2$ the external Alfv\'{e}n speed is smaller than the internal and no waves propagate. 
Lastly note, that in this investigation we assume that $\Beqzi \geq \Beqze$ and therefore $\rhoi = \rhoe$ has no admissible solution for $w$ when $\Beqzi = \Beqze$. 
Now, when $\Beqz$ is assumed to be constant, i.e. $\Beqzi = \Beqze = \Beqz$, using \eref{eqn:w:constant:b}, $\rho_A / \rhoi$, $B_{zA} / \Beqzi$ and $Z$ simplify to,
\begin{align}
\frac{\rho_A}{\rhoi} = \frac{1}{h^2}, && \frac{B_{zA}}{\Beqzi} = 1, && Z = |1 - \chi|, 
\end{align}
resulting in $\gamma_A$ (\eref{eqn:damping:factor:weak:twist})
\begin{align}
\gamma_A = \frac{\omega_0}{h^2}\frac{\pi}{|1 - \chi|} \frac{\ell}{r_e} \frac{B_{\varphi A}^2}{B_{zA}^2} {=} \frac{\omega_{Ai}}{h}\frac{\pi}{|1 - \chi|} \frac{\ell}{r_e} \frac{B_{\varphi A}^2}{B_{zA}^2}.
\end{align}

To obtain the damping time normalized by the period of the wave, we use a typical wavelength $\kwvZ = \pi / L$, where $L$ is the characteristic length of the tube, the associated period is $\tau = 2 L / h \vAi$ (see \eref{eqn:lwl:twist:first:order:simp}) the damping time ($1 / \gamma_A$) for modes in the continuum as a multiple of the wave period is, 
\begin{dmath}\label{eqn:damping:time:bi:d:be}
\tau_{d} = \frac{Z}{2 \pi^2} \frac{r_e}{\ell} \frac{\rhoi}{\rho_A} \frac{B_{zA}^2}{B_{\varphi A}^2} \tau, 
\end{dmath}
a contour map of this equation for $\zeta \in (0,1)$ and $\chi \in (0, \zeta^2/h^2)$, can be seen in \fref{fig:damping:time:bi:d:be}. When $\Beqzi = \Beqze$ the damping time becomes,
\begin{dmath}\label{eqn:damping:time:bi:e:be}
\tau_{d} = h^2 \frac{|1-\chi|}{2 \pi^2} \frac{r_e}{\ell} \frac{B_{zA}^2}{B_{\varphi A}^2}  \tau.
\end{dmath}
The long wavelength limit approximation of the damping rate $\gamma_{A}$ in \eref{eqn:damping:factor:bi:d:be}, is accurate to $\approx 10^{-6}$ at $\kwvZ r_e = 1$ when compared with the numerical solution of the dispersion relation in \eref{eqn:alfven:continuum:dr}. This accuracy is better than $10^{-6}$ for $\kwvZ r_e < 0.1$ and is calculated using the maximum of the root mean squared error (RMSE), 
\begin{dmath}
\mbox{RMS Error} =  \left( \frac{1}{N-1}\sum_{i=1}^{N} \left(\frac{\gamma_{A} - \hat{\gamma}_{A}}{\gamma_{A}}\right)^2 \right)^{1/2}.
\end{dmath}
In this equation, $\gamma_{A}$ is the numerically calculated damping rate, $\hat{\gamma}_{A}$ is the theoretical approximation in \eref{eqn:damping:factor:bi:d:be} and $N$ is the number of samples. For this error estimate we used $10^4$ samples in the parameter space ($\chi,\zeta,\ell/r_e,\Beqphi/\Beqz$), uniformly distributed\footnote{Since the parameter space is not a hypercube, e.g. see \fref{fig:damping:time:ratio}, we used rejection sampling for invalid parameter combinations until the desired number of samples was achieved.}.

Works investigating resonant absorption in the context of solar atmospheric conditions tend to consider solely a radial non-uniformity in either the magnetic field or density. However, accounting for radial variation in both the magnetic field and density can lead to significant variation in the estimated damping times. The ratio of \eref{eqn:damping:time:bi:d:be} over \eref{eqn:damping:time:bi:e:be} is, 
\begin{dmath}\label{eqn:damping:time:ratio}
\frac{\tau_{d}(\chi,\zeta)}{\tau_{d}(\chi,1)} = \frac{Z}{|1-\chi|} \frac{\rhoi}{\rho_A} \frac{1}{h^2}
\end{dmath}
and in \fref{fig:damping:time:ratio} a contour map is shown for $\zeta \in (0,1)$ and $\chi \in (0,\zeta^2 / h^2)$. 

It can be seen from \fref{fig:damping:time:bi:d:be} that the behavior of the damping rate with respect to changes in the density contrast is in some regions exactly the opposite to that for the kink mode \citep[][see for example]{Goossens:1992aa}. Namely, in a \textit{roughly} triangular region in \fref{fig:damping:time:bi:d:be} the damping rate is proportional to $\sim 1 / \chi$, in contrast to the kink mode where the damping rate is proportional to $\sim \chi$. Similar behavior has been been shown to exist in the leaky regime for sausage modes \citep{Vasheghani-Farahani:2014aa}. The factor in \eref{eqn:damping:time:bi:d:be} that determines this behavior is $Z \rhoi / \rho_A$. We approximate the local minimum in the $\chi$ direction by evaluating the partial derivative of $Z$ with respect to $\chi$,
\begin{dmath}
\pderiv{Z}{\chi} = \frac{-2 \zeta +h^2 (\chi -1)+2}{h \sqrt{4 (\zeta -1) (\zeta -\chi )+h^2 (\chi -1)^2}},
\end{dmath}
which is subsequently equated to $0$. From this we obtain a relation $\chi = (2/h^2) \zeta + b$ and $b$ is identified by noting that at $\zeta = 1$, the maximum value for $\chi$ is $1/h^2$, thus the approximation is, 
\begin{dmath}\label{eqn:qstat:approx}
\chi = \frac{2}{h^2} \zeta - \frac{1}{h^2}. 
\end{dmath}
As the remaining terms in \eref{eqn:damping:time:bi:d:be} do not vary with $\chi$ and $\zeta$ (note the ratio $\Beqphi/\Beqz$ is held fixed) this approximation holds for all valid parameters. This approximation allows us to estimate in which regime a specific parameter combination is. Namely, for parameter combinations that are below the line described by \eref{eqn:qstat:approx}, for increasing density contrast ($\chi \downarrow$), damping will be slower ($\tau_d \uparrow$). For parameter combinations that result in points above this line, increasing the density contrast ($\chi\downarrow$) results in decreasing damping time ($\tau_d \downarrow$), namely waves will decay faster. This is illustrated in \fref{fig:damping:time:bi:d:be} as a yellow line (\eqref{eqn:qstat:approx}) and the exact inflection points are marked with a green line.

Given the form of \eref{eqn:damping:time:bi:d:be}, and especially that of \eref{eqn:damping:time:bi:e:be}, a comparison with previous results for the kink mode is in order, particularly the expression for the damping rate obtained by \cite{goossens1992resonant} and later by \cite{ruderman2002damping}. In \citep{ruderman2002damping}, and Equation (73) in that work, using the notation in this work, reads as follows,
\begin{dmath}\label{eqn:damping:factor:ruderman}
\tau_{d} = \frac{2}{\pi} \frac{r_e}{\ell} \frac{1 - \chi}{1+\chi} \tau.
\end{dmath}
The relative magnitude of the damping time shown in \eref{eqn:damping:factor:ruderman} and \eref{eqn:damping:time:bi:e:be} is, 
\begin{dmath}
\frac{\tau_{d,Axisymmetric}}{\tau_{d,Kink}} = \frac{h^2}{4 \pi} \frac{(1-\chi)^2}{1+\chi} \frac{B_{zA}^2}{B_{\varphi A}^2}. 
\end{dmath}
It is evident that there exists a region in the parameter space of $(\chi, \Beqphi/\Beqz)$ for which $\tau_{d,Axisymmetric}$ is smaller when compared with $\tau_{d,Kink}$, however, this comparison is given here just as a reference and caution should be exercised in its interpretation since the damping, $\tau_{d,Kink}$ in \cite{ruderman2002damping} was calculated for the kink mode without magnetic twist. It is possible that magnetic twist amplifies dissipation in the kink mode, and therefore still, dissipation for the kink mode may be larger than that of axisymmetric modes. 

\subsection{Numerical Solution of Dispersion Relation in the Alfv\'{e}n Continuum}
\begin{figure}
\centering
\epsscale{1}
\plotone{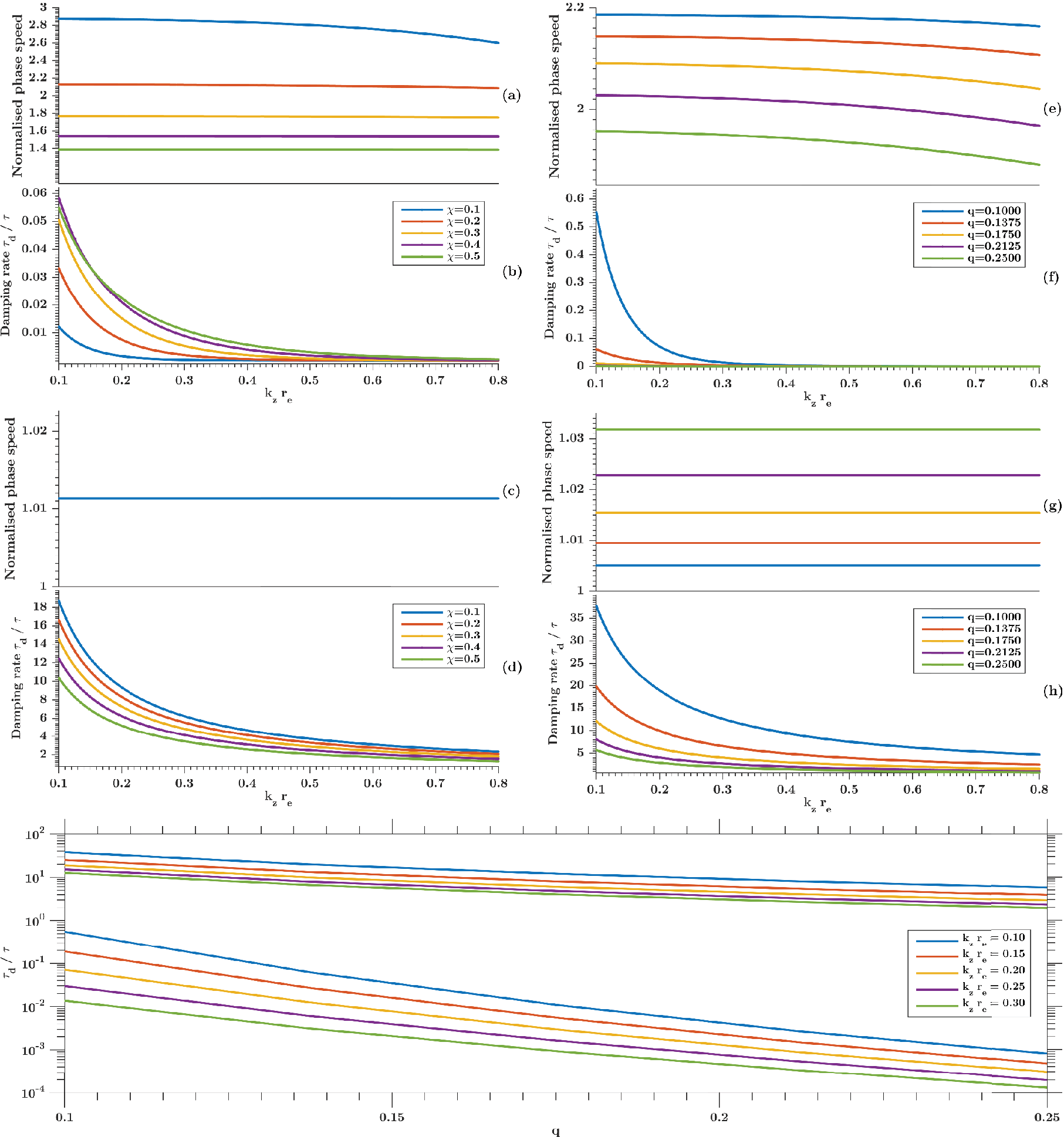}
\caption[]{Numerical solutions of the dispersion equation \eref{eqn:alfven:continuum:dr} for $\chi = \{0.1,0.2,0.3,0.4,0.5\}$, $q = 0.15$, $\zeta = 1$ and $\ell / r_e = 0.1$ for panels \textbf{(a)} to \textbf{(d)} ($\chi = \rhoe / \rhoi, \zeta = B_{ze} / B_{zi}, q = B_{\varphi A} / B_{zA}$) and $\chi = 0.2$, $q = \{0.1, 0.1375,0.175,0.2125,0.25\}$, $\zeta = 1$ and $\ell / r_e = 0.1$ for panels \textbf{(e)} to \textbf{(h)}. The panels \textbf{(a)}, \textbf{(c)}, \textbf{(e)} and \textbf{(g)} depict the normalized phase velocity and the panels \textbf{(b)}, \textbf{(d)}, \textbf{(f)} and \textbf{(h)} the corresponding normalized damping rates. The bottom panel shows a logarithmic plot of the damping time versus magnetic twist for different values of $\kwvZ r_e$. All solutions have been obtained numerically by solving \eref{eqn:alfven:continuum:dr}.\label{fig:alfven:cont:dispersion:soln}}
\end{figure}

We have solved \eref{eqn:alfven:continuum:dr} numerically, using $\omega_0$ obtained in \eref{eqn:lwl:twist:first:order} as an initial point in the solver. Additionally, by means of investigating whether another solution exists, we solved the dispersion relation again with a random $\omega_0$ in the range $(\vAi, \vAe)$. The solutions and their associated damping rates can be seen in \fref{fig:alfven:cont:dispersion:soln}. It is interesting that there exists another solution in the long wavelength limit that we could not obtain from our analysis in \sref{sec:long}. However, given that for this solution $\tau_d \ll \tau$, it is unlikely that this mode will be observed.

Now, it has been shown that the singularity about the resonance point at $r_A$ is logarithmic for $\Xirp$ ($ln(| r - r_A |)$) and $1 / (r - r_A)$ for $\Xiperp$, so the dynamics will be governed by $\Xiperp$ since $\Xirp / \Xiperp \rightarrow 0$ as $r \rightarrow r_A$ and therefore $\Xiperp \gg \Xirep$ in the neighbourhood of the resonant point \citep{Poedts:1989aa,Sakurai:1991aa}. Also the $\Xirp$ component provides its energy to the resonant layer \citep{goossens2011resonant} and therefore the characteristic expansion and contraction of axisymmetric modes will be reduced. These facts, along with the proximity of the solution corresponding to the long wavelength limit approximation to \sref{sec:long} to the internal Afv\'{e}n speed suggest that these waves would appear in observations to have properties similar to Alfv\'{e}n waves. Given that pure Alfv\'{e}n waves require $\nabla \cdot \Xip$ to be identically zero, and, a driving mechanism that is solely torsional, we argue that observed \textit{Alfv\'{e}n} waves are much more likely to be axisymmetric waves, as these do not have these strict requirements. In \fref{fig:alfven:cont:dispersion:soln} panels \textbf{(a)} through \textbf{(d)} show solutions for different values of $\chi$ while panels \textbf{(e)} through \textbf{(h)} solutions are shown when $q$ is allowed to vary. The damping time for the solution for which we have an analytical approximation (see panels \textbf{(c)}) and \textbf{(d)}) increases ($\tau_d \uparrow$) for increasing density contrast ($\chi \downarrow$) while the other solution exhibits the opposite behaviour (see panels \textbf{(a)} and \textbf{(b)}) namely $\tau_d \downarrow$ for $\chi \downarrow$. However, the damping time for both solutions decreases ($\tau_d \downarrow$) for increasing magnetic twist $q \uparrow$. The bottom panel of \fref{fig:alfven:cont:dispersion:soln} shows a different view of the damping times as a function of the magnetic tiwst ($q$) shown in panels (f) and (h) at $\kwvZ r_e = \{0.1,0.15,0.2,0.25,0.3\}$. From this view it can be seen that the solutions in (e) are much more sensitive to variations in the mangetic twist when compared with the solutions in panel (g). This sensitivity, in combination with the fact that for extremely small twist the sausage cut-off is reintroduced \citep{Giagkiozis:2015apj}, means that this mode will be observable for a very small interval of magnetic twist. The mode shown in panels (c) and (d) does not present this difficulty, and therefore we expect that observation of this mode is more likely. In both cases the solution corresponding to the analytic approximation remains very close to the internal Alfv\'{e}n speed which is equal to $1$ in \fref{fig:alfven:cont:dispersion:soln}. Since $\omega_0$ from \eref{eqn:lwl:twist:first:order:simp} depends on $q$, $B_{zA}$ and the internal sound speed, these modes will appear to have a strong Alfv\'{e}n character for virtually all valid parameter combinations. Lastly, $\kwvR$ can be likened to the wavenumber in the radial direction, and, since in the long wavelength limit $\kwvR$ is proportional to $\kwvZ$, as $\kwvZ$ increases, the wavelength in the radial direction decreases and couples with the thin inhomogeneous layer more closely and therefore more energy per wavelength is absorbed and thus the damping time is reduced (see \fref{fig:alfven:cont:dispersion:soln}). 

\section{Connection to Observations}\label{sec:con:observations}
\begin{figure}
\centering
\epsscale{1}
\plotone{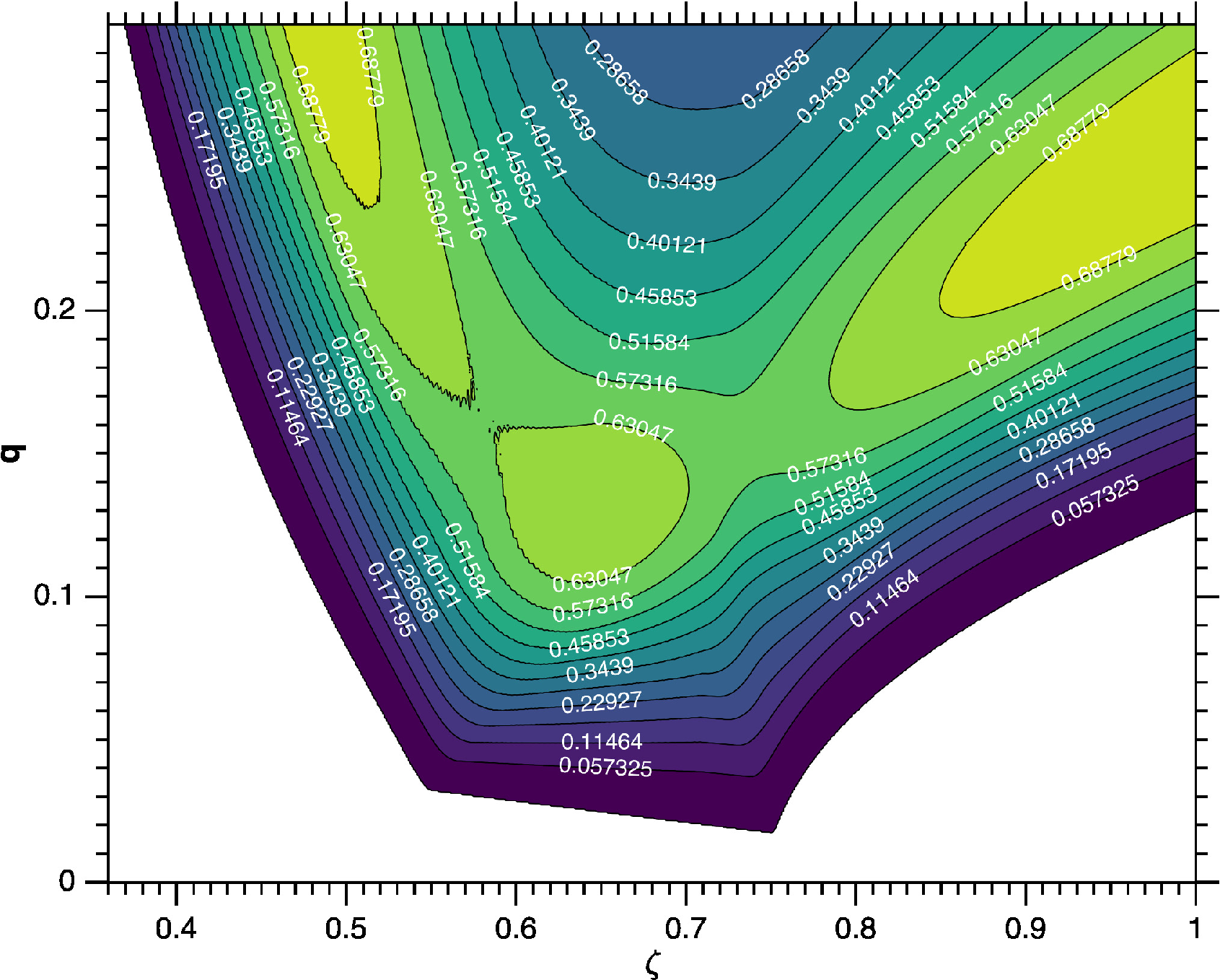}
\caption[]{Contour map of the estimated probability (see \eref{eqn:observation:probability}) that an axisymmetric mode can be observed to have a normalized damping time $\bar{\tau}_d$ in the range $(1,3)$, for a given combination of $(\zeta,q)$, i.e. magnetic field contrast and twist respectively. The free parameters are $\zeta \in (0.35,1)$ and $q=B_{\varphi A}/B_{zA} \in (0,0.3)$ and the integration parameters are $\chi \in (0.5,1)$ and $\ell/r_e \in (0.1, 0.5)$. The white region represents $0$ probability. \label{fig:damping:1:to:3:BQ}}
\end{figure}
\begin{figure}
\centering
\epsscale{1}
\plotone{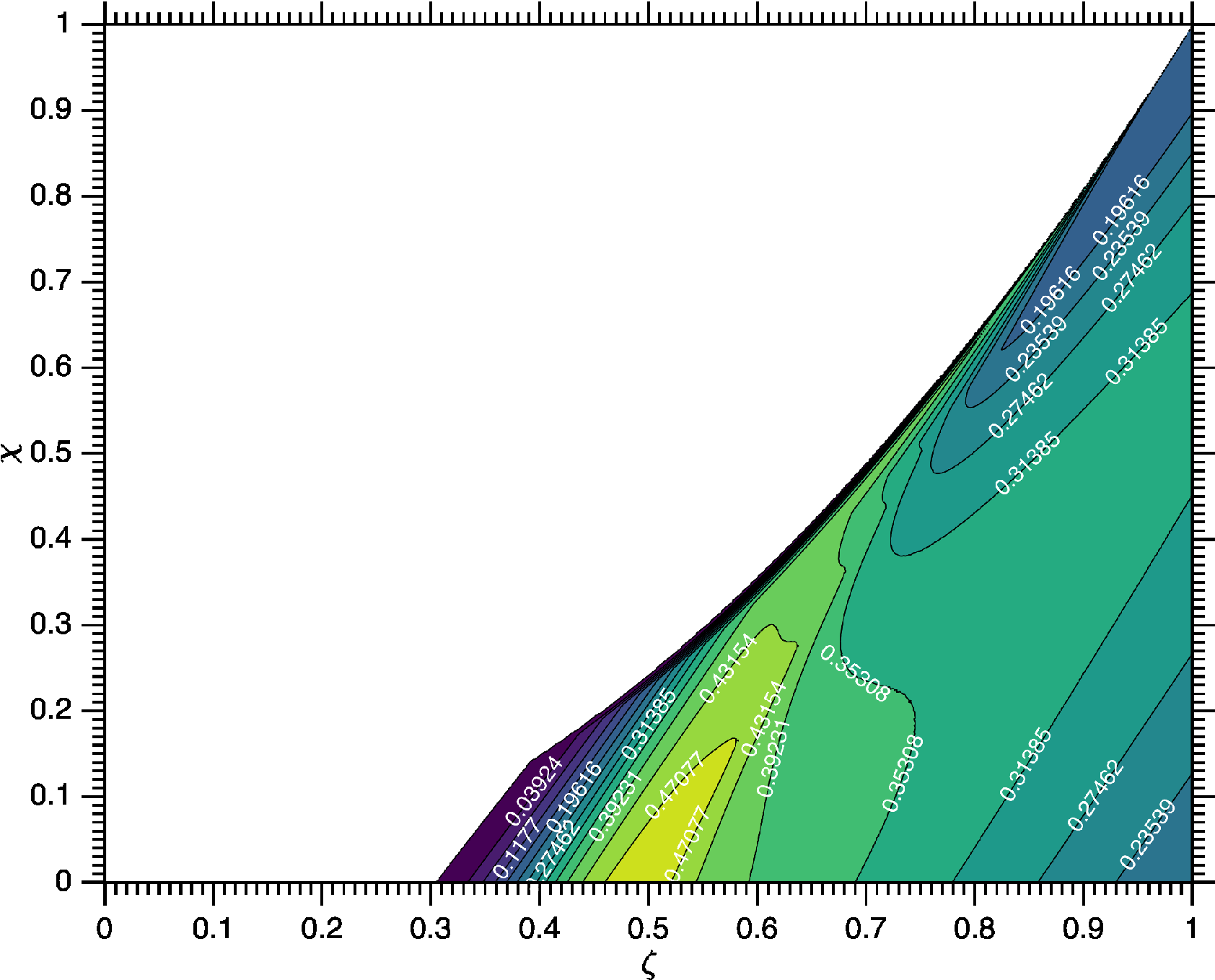}
\caption[]{Contour map of the estimated probability for an axisymmetric mode to be observed to have a normalized damping time $\bar{\tau}_d$ in the range $(1,3)$ for a point in $(\zeta,\chi)$, i.e. magnetic field and density contrast respectively. The free parameters are $\zeta \in (0,1)$ and $\chi \in (0,1)$ and the integration parameters are $q=B_{\varphi A}/B_{zA} \in (0,0.3)$ and $\ell/r_e \in (0.1, 0.5)$. Similarly to \fref{fig:damping:1:to:3:BQ} the white region in this map represents an estimated probability of $0$ of observing resonantly absorbed axisymmetric modes for the particular set of parameter combinations. \label{fig:damping:1:to:3:BChi}}
\end{figure}
\notetoeditor{ {\fref{fig:damping:1:to:3:BQ}} and \fref{fig:damping:1:to:3:BChi} should appear side-by-side in the text.}

Reports of observations of axisymmetric modes (sausage modes) are increasing in frequency in the recent literature. For example quasi-periodic pulsations in solar flares are believed to be associated with the kink and sausage mode \citep[see for example][]{Van-Doorsselaere:2011aa,Nakariakov:2011aa,Nakariakov:2012aa,De-Moortel:2012aa,Kolotkov:2015aa}. Even more interestingly some of these pulsations appear to have periods in the interval $(15,100)$ seconds which could be consistent with the results in the present investigation if the length-scale of these pulsations is on the same order as the length-scale of coronal flux tubes $\approx 100Mm$. Furthermore the results by \cite{morton2012observations} suggest that axisymmetric modes are ubiquitous and that they appear to be coexistent with kink modes. This coexistence further supports the argument by \cite{Arregui:2015aa,Arregui:2015ab,Arregui:2015ac} that Bayesian analysis is an essential tool for the identification of the \textit{likely} wave modes present in observations as well as a more systematic way for the appropriate model selection. The uncertainty in determining the parameters for the kink mode led \cite{Verwichte:2013ab} to perform a statistical analysis as a way to narrow the range of their values. This departure from \textit{certainty} and convergence towards probabilistic inference models for solar observations is, in our view, long overdue. 

However, despite this increase in interest in axisymmetric modes, the relation that approximates their expected damping rate, see \eref{eqn:damping:time:bi:d:be}, requires knowledge of four parameters. Namely, the density and magnetic field contrast, the relative magnetic twist and the ratio of the thickness of the inhomogeneous layer versus the tube radius, i.e. $(\chi,\zeta,q=B_{\varphi A}/B_{zA},\ell/r_e)$. In contrast to the large body of observational evidence for the kink mode, observations of sausage waves are relatively scarce. This makes impossible an analysis similar to \cite{Verwichte:2013ab} for these modes. Therefore, we adopt a different approach, a probabilistic approach which is related to the use of Bayesian inference suggested by \cite{Arregui:2015aa}.

As a first step towards improving this situation we provide a way to estimate the probability that an observed sausage wave has a damping rate within a specified range, given that, one or more of the four parameters in \eref{eqn:damping:factor:bi:d:be} are known. The assumptions required for the validity of this estimate are the following: 
\begin{itemize}
\item The four parameters in \eref{eqn:damping:factor:bi:d:be} are independent, i.e. no parameter is a function of the others.
\item The likelihood of any combination in the parameter space is the same. That is to say that there exists no preferred combination of parameters.
\end{itemize}
These assumptions are difficult to prove, especially given that there exist no statistical analyses of the properties of sausage waves and reliable estimates of all four parameters. Since we do not know if there is, in fact, a set of preferred parameters, these assumptions are required for an unbiased estimate. Acknowledging these uncertainties, we make a first attempt in identifying the probability predicted by our model that a wave with the characteristics described in this investigation is resonantly damped in the long wavelength limit with a damping rate given by \eref{eqn:damping:time:bi:d:be} for a given parameter combination. 

The aforementioned probability can be estimated as follows. First, we identify the parameters for which reasonably good estimates are available. These parameters we refer to as \textit{free} parameters denoted by $f$. The remaining parameters we refer to as \textit{integration} parameters and are denoted by $i$. Subsequently, a domain is defined for the integration parameters. Then the probability of the damping rate being within the open interval $(a,b)$ is given by, 
\begin{align}\label{eqn:observation:probability}
P(a,b;f_1,\dots,f_n) &= \frac{ \int_{C} d i_1 \dots d i_{4-n} w(i_1,\dots,i_{4-n}) I_{\bar{\tau}_d>a,\bar{\tau}_d<b}\left[ \cdot \right] }{ \int_{C} d i_1 \dots d i_{4-n} w(i_1,\dots,i_{4-n}) I_1\left[ \cdot \right]  }, \\
I_{\bar{\tau}_d>a,\bar{\tau}_d<b}\left[ \cdot \right] &= I_{\bar{\tau}_d>a,\bar{\tau}_d<b}\left[ \bar{\tau}_d(i_1,\dots,i_{4-n};f_1,\dots,f_n) \right], \\
I_1\left[ \cdot \right] &= I_1\left[ \bar{\tau}_d(i_1,\dots,i_{4-n};f_1,\dots,f_n) \right],
\end{align}
where $C$ is the domain of integration defined as the set of all elements in the integration parameter space that are valid according to the analysis in this work, and, $\bar{\tau}_d = \tau_d / \tau$. The function $I(\cdot)$ is an indicator function and $n = \{1,2,3\}$, i.e. an estimate for at least one parameter is necessary. When the indicator function is subscripted with $1$ it simply returns one when the parameter combination is valid, namely the integral in the denominator of \eref{eqn:observation:probability} simply returns the area where $\bar{\tau}_d$ is defined. The indicator function in the numerator is defined as follows, 
\begin{dmath}
I_{\bar{\tau}_d>a,\bar{\tau}_d<b}\left[ \bar{\tau}_d(i_1,\dots,i_{4-n};f_1,\dots,f_n) \right] = \twopartdefc{\displaystyle  1 } { a<\bar{\tau}_d<b,} {\displaystyle  0 } {\mbox{otherwise},}
\end{dmath}
namely this function returns $1$ when the normalized damping rate ($\bar{\tau}_d$) is within the open interval $(a,b)$, and therefore the numerator of \eref{eqn:observation:probability} returns the area in $C$ for with the normalized damping rate is within the interval $(a,b)$. The function $w(i_1,i_2)$ is a weighting function that is non-negative, and, its integral over $C$ is equal to $1$. As we have assumed that every combination in the integration parameter space $(i_1,i_2)$ is equally likely, this function is simply a constant and simplifies out from the integrals. The effect of this function is similar to the prior information in Bayesian inference. Therefore if relevant information of a specific preference in parameter space is present in the solar atmosphere, this can be taken into account by appropriately modifying $w(\cdot)$. 

For the contour maps \fref{fig:damping:1:to:3:BQ} and \fref{fig:damping:1:to:3:BChi} it is assumed that the free parameters in \eref{eqn:observation:probability} are $(\zeta,q)$ and $(\zeta,\chi)$ respectively. For this case \eref{eqn:observation:probability} becomes
\begin{dmath}\label{eqn:observation:probability:2}
P(a,b;f_1,f_2) = \frac{ \int_{C} d i_1 d i_2 w(i_1,i_2) I_{\bar{\tau}_d>a,\bar{\tau}_d<b}\left[ \bar{\tau}_d(i_1,i_2;f_1,f_2) \right] }{ \int_{C} d i_1 d i_2 w(i_1,i_2) I_1\left[ \bar{\tau}_d(i_1,i_2;f_1,f_2) \right]  }.
\end{dmath}
Our rationale for selecting the limits for the integration parameters in \fref{fig:damping:1:to:3:BQ} is based on the reported values for the parameters $(\chi,\ell/r_e)$ and normalized damping rate, in \cite{Aschwanden:2003aa} for the kink mode. These authors used $11$ cases of observed damping kink oscillations and their estimates for these parameters are as follows: $\chi \approx 0.1$, $\ell/r_e \in (0.1,0.5)$ and the observed normalized damping rates were in the interval $(1,3)$. The $q$ parameter has been selected in an interval that ensures that the magnetic twist is small. As can be seen in both \fref{fig:damping:1:to:3:BQ} and \fref{fig:damping:1:to:3:BChi} the probability for the resonantly absorbed axisymmetric modes for a wide range in parameters is significantly high. Also, as seen in \fref{fig:damping:1:to:3:BQ} normalised damping times in the interval $(1,3)$ are possible even for extremely small magnetic twist ($\approx 0.02$).

\begin{figure}
\centering
\plotone{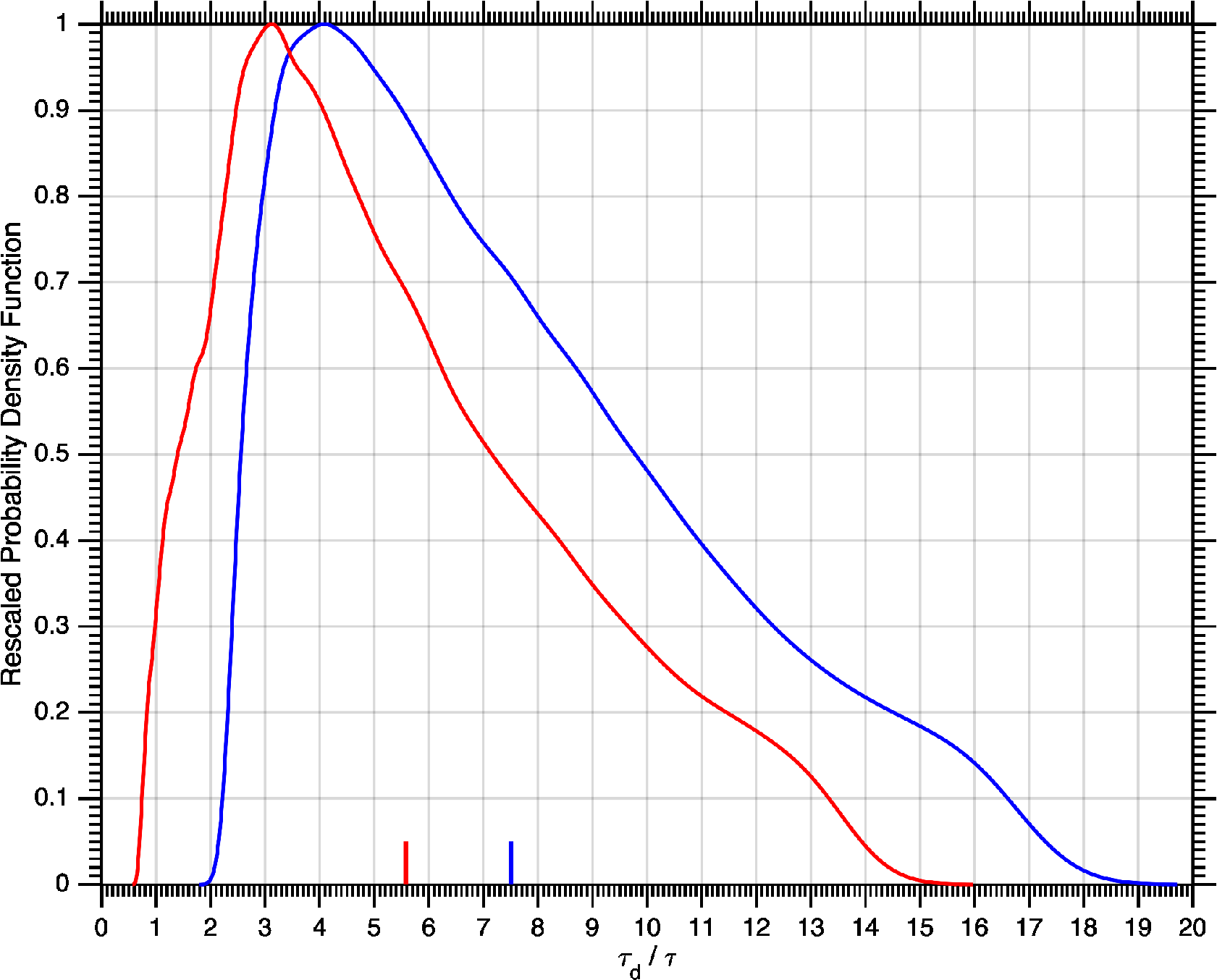}
\caption[]{Rescaled probability density functions (PDF) of the normalized damping time using parameter estimates from \cite{morton2012observations} (\textbf{blue}) and \cite{Van-Doorsselaere:2011aa} (\textbf{red}). For illustration purposes the scaling in both PDFs is such that their maximum is equal to $1$. The support for the \textit{blue} PDF is $(1.79,19.72)$ and the expected value for the damping time is $E[\tau_d/\tau] = 7.49$. Similarly the support for the \textit{red} PDF is $(0.56,15.97)$ with an expected value for the damping time $E[\tau_d/\tau] = 5.58$. The intervals used for the parameters ($\chi = \rhoe / \rhoi, \zeta = B_{ze} / B_{zi}, q = B_{\varphi A} / B_{zA}$) and the associated assumptions are detailed in \sref{sec:con:observations}. \label{fig:PDFs}}
\end{figure}

In the case where more information is available Monte Carlo simulation can be used to estimate the probability density function (PDF) of the normalized damping time. We illustrate this with two examples. First we use the estimates from \cite{morton2012observations}. In that work the magnetic field is assumed to be constant inside and outside the flux tube but this assumption is unlikely to be identically satisfied so we allow a small variation in $\zeta$ in the interval $\zeta \in (0.95, 1)$. The density contrast is taken to be in the interval $(10^{-2}, 10^{-1})$. Since \cite{morton2012observations} do not provide an estimate for the width of the inhomogenous layer we allow it to vary uniformly in $(0.1, 0.5)$, an interval that is in line with estimates in \cite{goossens2002coronal} and \cite{Aschwanden:2003aa}. In both examples we assume the magnetic twist is within the interval $(0.1,0.2)$. Using these intervals and assuming a uniform distribution we sample \eref{eqn:damping:time:bi:d:be} $10^{6}$ times. The estimated PDF for this set of parameters is the blue curve in \fref{fig:PDFs}. The blue vertical line is the expectation value which is equal to $E[\tau_d / \tau] = 7.49$.

In the second example we use parameter estimates from \cite{Van-Doorsselaere:2011aa}. Assuming a $H$ plasma, $\rho = N m_p$, $p = N k_B T$ where $N$ is the number density, $m_p$ the proton mass and $k_B$ is the Boltzmann constant. With the plasma-$\beta$ equal to $\beta = 2 \mu_0 p / B^2$ and the assumption that $\beta_e \ll \beta_i$ \citep{Van-Doorsselaere:2011aa} we obtain
\begin{dmath}\label{eqn:beta:assumption:vand}
\frac{\beta_i}{\beta_e} {=} \frac{\zeta^2}{\chi} \frac{T_i}{T_e} \gg 1.
\end{dmath}
Assuming a lower limit for $\beta_i / \beta_e \geq 100$ and a hot flux tube, $T_i/T_e = 10$, we can restrict $\zeta$ and $\chi$ to  
\begin{align}
\frac{1}{200} &\leq \chi \leq \frac{\beta_e}{\beta_i} \frac{T_i}{T_e}, \\
\left( \chi \frac{T_e}{T_i} \frac{\beta_i}{\beta_e}  \right)^{1/2} &\leq \zeta \leq 1.
\end{align}
The lower limit for $\chi$ is considered as a minimum contrast in \cite{Van-Doorsselaere:2011aa} to avoid the sausage cut-off. However, in the presence of very weak magnetic twist this cut-off is removed, and therefore we don't need to assume extreme values for the density contrast. The upper limit for $\chi$ and lower limit for $\zeta$ are taken so that \eref{eqn:beta:assumption:vand} is satisfied. The resulting PDF can be seen in \fref{fig:PDFs}. It is interesting that the expected value for the damping time in this case is $5.58$ which is very close to observed damping ($\tau_d / \tau = 6$) of a mode that is believed to be a fast sausage mode \citep{Kolotkov:2015aa}. It is apparent from \fref{fig:PDFs} that the PDFs cannot be approximated well using a normal distribution, and therefore their use for obtaining estimates of the damping time from results like \eref{eqn:damping:time:bi:d:be} in this work, and, similar equations \citep[e.g.][]{goossens1992resonant,ruderman2002damping} can be misleading. In contrast, Monte Carlo simulation and non-parametric density estimation can be quite useful tools for exploring this type of problems.

\section{Discussion and Conclusions} \label{sec:conclusions}
Theoretically, it has been known for some time, that, in the presence of weak magnetic twist, axisymmetric modes will be resonantly damped \citep[see for example][]{goossens1992resonant}. In this work we have calculated, for the first time, a dispersion relation for resonantly damped axisymmetric modes, in the spectrum of the Alfv\'{e}n continuum and derived an approximation of the damping time in the long wavelength limit. We have shown that the damping time can be comparable to that observed for the kink mode in the case that there is no magnetic twist. Furthermore, we solved the resulting equation (see \eref{eqn:alfven:continuum:dr} and \eref{eqn:damping:factor:bi:d:be}) analytically and, i) we confirmed the validity of our approximation, and, ii) we found an additional solution that decays much faster in comparison. The resulting approximation in the long wavelength limit shows that the damping time is proportional to the magnetic twist and inversely proportional to the density contrast. It is interesting to note that \cite{Vasheghani-Farahani:2014aa} who investigated the damping of fast sausage modes in the leaky regime found a similar relation between the damping time and the density contrast. However, in that work a very large density contrast is required to allow observation of the sausage mode. This is not the case for one of the results of this work, which even for modest density contrasts (see \fref{fig:damping:1:to:3:BQ} and \fref{fig:damping:1:to:3:BChi}) the damping time is within one to three periods of the wave.

Of the two solutions that we have uncovered, only the one with a phase velocity close the internal Alfv\'{e}n speed has, for some parameter combinations, damping times that would allow observation. The other solution is found to be damped on time scales $\approx 10^{-2}-10^{-1}$ of the wave period as seen in \fref{fig:alfven:cont:dispersion:soln}, which means that its observation would be extremely challenging. On the other hand although the predicted damping times for the solution whose phase speed is close to the Alfv\'{e}n speed are large enough to allow observation. Also, the fact that its phase speed is so close the internal Alfv\'{e}n speed along with the dominance of the $\Xiperp$ component in the wave dynamics means that the character of this wave will be predominantly Alfv\'{e}nic \citep{goossens2011resonant}. Because of this, we argue that it is possible that resonantly damped sausage waves have already been observed, albeit in the guise of Aflv\'{e}n waves, see for example \cite{jess2009alfven}. 

Lastly, we estimated the damping time for the parameters presented by \cite{morton2012observations} and \cite{Van-Doorsselaere:2011aa} and interestingly the expected damping time is very close to the observed damping in quasi-periodic pulsations by \cite{Kolotkov:2015aa} that are believed to be fast sausage waves. We find, subject to certain assumptions, that axisymmetric modes appear to be quite important conduits for energy transfer in the solar atmosphere. Perhaps even more important than pure Alfv\'{e}n waves, given that the excitation mechanism for sausage modes in weakly twisted magnetic flux tubes appears to be more readily available in comparison to the required purely torsional drivers for Alfv\'{e}n waves \citep{Giagkiozis:2015apj}.

\acknowledgments
I.G. would like to acknowledge the Faculty of Science of the University of Sheffield for the SHINE studentship. I.G. also thanks T.V.D. for financial support during his visit to KU Leuven. M.G. is grateful to Belspo's IAP P7/08 CHARM and KU Leuven GOA-2015-014. G.V. and V.F. would like to acknowledge the STFC for funding received (Grant number ST/M000826/1). T.V.D. has received funding from the Odysseus programme of the FWO-Vlaanderen, and would also like to acknowledge the framework of Belspo's IAP P7/08 CHARM and the GOA-2015-014 of the Research Council of the KU Leuven.

\end{document}